\documentclass[aps,twocolumn,prd,showpacs,amsmath,amssymb,superscriptaddress,sort&compress,nofootinbib,preprintnumbers]{revtex4-1}

\usepackage{graphicx}
\usepackage{xspace}
\usepackage{bm}
\usepackage{tikz}
\usepackage[colorlinks=true, citecolor=blue, urlcolor=blue, filecolor=blue]{hyperref}
\usepackage{enumitem}
\usepackage{afterpage}

\newcommand{\aachen}{Institute for Theoretical Particle Physics and Cosmology (TTK), RWTH Aachen University, D-52056 Aachen, Germany}
\newcommand{\queens}{Department of Physics, Engineering Physics and Astronomy, Queen's University, Kingston ON K7L 3N6, Canada}
\newcommand{\imperial}{Department of Physics, Imperial College London, Blackett Laboratory, Prince Consort Road, London SW7 2AZ, UK}
\newcommand{\cambridge}{Cavendish Laboratory, University of Cambridge, JJ Thomson Avenue, Cambridge, CB3 0HE, UK}
\newcommand{\oslo}{Department of Physics, University of Oslo, Box 1048, Blindern, N-0316 Oslo, Norway}
\newcommand{\adelaide}{ARC Centre for Dark Matter Particle Physics, Department of Physics, University of Adelaide, Adelaide, SA 5005, Australia}
\newcommand{\monash}{School of Physics and Astronomy, Monash University, Melbourne, VIC 3800, Australia}
\newcommand{\mcdonald}{Arthur B. McDonald Canadian Astroparticle Physics Research Institute, Kingston ON K7L 3N6, Canada}
\newcommand{\okc}{Oskar Klein Centre for Cosmoparticle Physics, AlbaNova University Centre, SE-10691 Stockholm, Sweden}
\newcommand{\perimeter}{Perimeter Institute for Theoretical Physics, Waterloo ON N2L 2Y5, Canada}
\newcommand{\uq}{School of Mathematics and Physics, The University of Queensland, St.\ Lucia, Brisbane, QLD 4072, Australia}

\newcommand{\kicc}{Kavli Institute for Cosmology, University of Cambridge, Madingley Road, Cambridge, CB3 0HA, UK}
\newcommand{\caius}{Gonville \& Caius College, Trinity Street, Cambridge, CB2 1TA, UK}

\newcommand{\gambit}{\GB}
\newcommand{\GB}{\textsf{GAMBIT}\xspace}
\newcommand{\cosmobit}{\textsf{CosmoBit}\xspace}
\newcommand{\precisionbit}{\textsf{PrecisionBit}\xspace}
\newcommand{\neutrinobit}{\textsf{NeutrinoBit}\xspace}
\newcommand{\scannerbit}{\textsf{ScannerBit}\xspace}
\newcommand{\Planck}{\emph{Planck}\xspace}
\newcommand{\nufit}{\textsf{NuFit}\xspace}
\newcommand{\lcdm}{$\Lambda$CDM\xspace}
\newcommand{\polychord}{\textsf{PolyChord}\xspace}
\newcommand{\montepython}{\textsf{MontePython}\xspace}
\newcommand{\alterbbn}{\textsf{AlterBBN}\xspace}
\newcommand{\class}{\textsf{CLASS}\xspace}

\newcommand{\mnuzero}{m_{\nu_0}}

\def\epjc{Eur.\ Phys.\ J.\ C}

\def\njp{New J.\ Phys.}

\def\arnps{Ann.~Rev.~Nucl.~\& Part.~Sci.}%
\def\apj{ApJ}%
\def\aap{A\&A}%
\def\jhep{JHEP}%
\def\jcap{JCAP}%
\def\mnras{MNRAS}%
\def\prd{Phys.\ Rev.\ D}%
\def\prl{Phys.\ Rev.\ Lett.}%
\def\plb{Phys. Lett. B}%
\def\epjp{Eur.~Phys.~J.~Plus}%
\def\epjc{Eur.~Phys.~J.~C}%
\def\njp{New J.~Phys.}

\begin{document}

\title{Strengthening the bound on the mass of the lightest neutrino \\ with terrestrial and cosmological experiments}

\author{The GAMBIT Cosmology Workgroup: Patrick St\"ocker}
\email{stoecker@physik.rwth-aachen.de}
\affiliation{\aachen}

\author{Csaba Bal{\'a}zs}
\affiliation{\monash}

\author{Sanjay Bloor}
\affiliation{\uq}
\affiliation{\imperial}

\author{Torsten Bringmann}
\affiliation{\oslo}

\author{Tom\'as E. Gonzalo}
\affiliation{\monash}

\author{Will Handley}
\affiliation{\cambridge}
\affiliation{\kicc}
\affiliation{\caius}

\author{Selim Hotinli}
\affiliation{\imperial}

\author{Cullan Howlett}
\email{c.howlett@uq.edu.au}
\affiliation{\uq}

\author{Felix Kahlhoefer}
\affiliation{\aachen}

\author{Janina J. Renk}
\email{janina.renk@fysik.su.se}
\affiliation{\uq}
\affiliation{\imperial}
\affiliation{\okc}

\author{Pat Scott}
\email{pat.scott@uq.edu.au}
\affiliation{\uq}
\affiliation{\imperial}

\author{Aaron C. Vincent}
\affiliation{\queens}
\affiliation{\mcdonald}
\affiliation{\perimeter}

\author{Martin White}
\affiliation{\adelaide}

\date{\today}

\begin{abstract}
\noindent We determine the upper limit on the mass of the lightest neutrino from the most robust recent cosmological and terrestrial data. Marginalizing over possible effective relativistic degrees of freedom at early times ($N_\mathrm{eff}$) and assuming normal mass ordering, the mass of the lightest neutrino is less than 0.037\,eV at 95\% confidence; with inverted ordering, the bound is 0.042\,eV. These results improve upon the strength and robustness of other recent limits and constrain the mass of the lightest neutrino to be barely larger than the largest mass splitting. We show the impacts of realistic mass models, and different sources of $N_\mathrm{eff}$.
\end{abstract}


\preprint{TTK-20-28, gambit-physics-2020}

\maketitle

\section{Introduction}

Neutrino masses are arguably the most concrete evidence to date of physics beyond the Standard Model (SM). Measurements of their flavor oscillations at reactor \cite{Gando:2013nba,An:2016srz,Adey:2018zwh,dchooz,Bak:2018ydk}, accelerator \cite{Adamson:2013whj,Adamson:2013ue,t2k,t2k2,sanchez_mayly_2018_1286758,Acero:2019ksn}, solar \cite{Cleveland:1998nv,Kaether:2010ag,Abdurashitov:2009tn,Aharmim:2011vm,Hosaka:2005um,Cravens:2008aa,Abe:2010hy,skiv,Bellini:2011rx,Bellini:2008mr,Bellini:2014uqa} and atmospheric \cite{Aartsen:2014yll,Abe:2017aap} experiments show that at least two of the three SM neutrinos must be massive.  While oscillation experiments probe mass-squared differences between eigenstates, the expansion history of the Universe, growth of cosmic structure and the cosmic microwave background (CMB) are sensitive to absolute masses, which determine when a neutrino becomes nonrelativistic; for most cosmological applications, this is expressed in terms of the sum of masses $\sum m_\nu$. Robust and precise inference on the mass of the lightest state can therefore only be obtained by combining the latest results of all these probes self-consistently, including associated uncertainties from each, as well as constraints on other relevant parameters from e.g.\ big bang nucleosynthesis (BBN) and late-time cosmological observables \cite{Wong:2011ip,Lesgourgues:2014zoa,Vagnozzi:2017ovm,Aghanim:2018eyx,Loureiro:2018pdz,Ivanov:2019hqk,Archidiacono:2020dvx}.

As no probe has yet directly measured the mass of a single neutrino,  the most convenient three-flavor parametrisation is in terms of the mass $\mnuzero$ of the lightest neutrino and two squared mass splittings, $\Delta m_{21}^2 \equiv m_2^2 - m_1^2$ and $\Delta m_{3l}^2 \equiv m_3^2 - m_l^2$. Here 1, 2 and 3 label the mass eigenstates with the largest component of $\nu_e$, $\nu_\mu$ and $\nu_\tau$, respectively.  Two mass orderings are presently permitted by the data: the normal ($m_1 < m_2 \ll m_3$; NH) and inverted hierarchies ($m_3 \ll m_1 < m_2$; IH).  $\Delta m_{3l}^2$ refers to the splitting between the lightest and heaviest states, i.e.\ $l=1$ for the NH and $l=2$ for the IH.  In terms of the splitting parameters, the physical masses are
\begin{eqnarray}
\mathrm{NH:}\ &(m_1^2,m_2^2,m_3^2) =& \mnuzero^2 + (0, \Delta m_{21}^2, \Delta m_{3l}^2)\\
\mathrm{IH:}\ &(m_3^2,m_1^2,m_2^2) =& \mnuzero^2 + (0, |\Delta m_{3l}^2|-\Delta m_{21}^2, |\Delta m_{3l}^2|).\nonumber
\end{eqnarray}

In this article, we make use of the new cosmology module \cosmobit \cite{cosmobit} within the beyond-the-SM global fitting package \GB \cite{gambit} in order to perform the most precise and robust combination to date of cosmological and experimental constraints on the mass of the lightest neutrino.  We include the most recent CMB likelihoods from \Planck \cite{Aghanim:2019ame}, recent three-flavor neutrino global fit results from \nufit \cite{Esteban:2018azc}, and correlated measurements of the baryon acoustic oscillation (BAO) scale by 6dF \cite{2011MNRAS.416.3017B}, SDSS-MGS \cite{2015MNRAS.449..835R}, BOSS DR12 \cite{Alam:2016hwk}, eBOSS DR14 \cite{Ata:2017dya,Bautista_2018,Blomqvist2019} and DES \cite{Abbott_2018}. We also compute and propagate the primordial helium abundance and number of effective relativistic degrees of freedom $N_\mathrm{eff}$ through all our calculations and likelihoods self-consistently and account for the uncertainty on the lifetime of the neutron. When computing bounds on neutrino masses, we illustrate the impact of different physical assumptions about $N_\mathrm{eff}$ and show the impact of the choice of neutrino mass model on the derived value of the Hubble parameter $H_0$, of particular interest given the present tension between expansion measurements at early and late times \cite{Riess:2019cxk,2019arXiv190704869W,2019ApJ...882...34F}.

\section{Methodology}

Our likelihoods are based on the latest and most constraining data implemented in \cosmobit \cite{cosmobit}, \neutrinobit \cite{RHN} and \precisionbit \cite{SDPBit}:
\begin{itemize}[noitemsep,topsep=0pt]
\item[(i)]Neutrino oscillations: Two-dimensional NH and IH $\Delta \chi^2$ tables for $\Delta m_{3l}^2$ and $\Delta m_{21}^2$ from \nufit \textsf{4.1} \cite{Esteban:2018azc}. These come from fits to the data from solar (Homestake chlorine \cite{Cleveland:1998nv}, Gallex/GNO \cite{Kaether:2010ag}, SAGE \cite{Abdurashitov:2009tn}, SNO \cite{Aharmim:2011vm}, four phases of Super-Kamiokande \cite{Hosaka:2005um,Cravens:2008aa,Abe:2010hy,skiv}, two phases of Borexino \cite{Bellini:2011rx,Bellini:2008mr,Bellini:2014uqa}), atmospheric (IceCube/DeepCore \cite{Aartsen:2014yll}, Super-Kamiokande \cite{Abe:2017aap}), reactor (KamLAND \cite{Gando:2013nba}, Double Chooz \cite{dchooz}, Daya Bay \cite{An:2016srz,Adey:2018zwh}, Reno \cite{Bak:2018ydk}), and accelerator experiments (MINOS \cite{Adamson:2013whj,Adamson:2013ue}, T2K \cite{t2k,t2k2}, NO$\nu$A \cite{sanchez_mayly_2018_1286758,Acero:2019ksn}). The other oscillation parameters (mixing angles $\theta_{ij}$ and \textit{CP} phase $\delta_{CP}$) have no bearing on neutrino mass studies and do not enter our analysis. Updates contained in \nufit \textsf{5.0} \cite{Esteban:2020cvm} released when this article was in the final stages of preparation include only small improvements to the likelihoods for $\Delta m_{3l}^2$ and $\Delta m_{21}^2$, so they have minimal impact on the results we show here.
\item[(ii)]BBN: Primordial abundances of $^4$He, $Y_\mathrm{p} = 0.245 \pm 0.003$ \cite{Tanabashi:2018oca} and deuterium, ${\rm D/H} = (2.527 \pm 0.030) \times 10^{-5}$ \cite{Cooke:2017cwo}.
\item[(iii)]CMB: \Planck 2018 baseline likelihoods consisting of high-$\ell$ and low-$\ell$ temperature and polarization data, plus CMB lensing \cite{Aghanim:2019ame}.
\item[(iv)]Supernovae Type Ia (SN Ia): $1048$ SN Ia included in the Pantheon compilation \cite{Scolnic:2017caz}.
\item[(v)]BAO scale: Measurements of the transverse comoving distance $D_{M}$ and the Hubble parameter $H(z)$ from the BOSS DR12 anisotropic consensus \cite{Alam:2016hwk}, $D_{M}$ from DES Y1 \cite{Abbott_2018}, and the volume-averaged distance $D_{V}$ from the combined 6dF and MGS galaxy surveys \cite{2011MNRAS.416.3017B, 2015MNRAS.449..835R, Carter_2018} and the eBOSS DR14 lumious red galaxy (LRG) and quasi-stellar objects (QSO) samples \cite{Ata:2017dya, Bautista_2018}. All measurements are relative to $r_{s}$, the radius of the sound horizon at the baryon-drag epoch.
\end{itemize}

We carefully consider correlations between overlapping samples in the BAO scale measurements. The 6dF+MGS result can be considered independent of the others, as these samples do not overlap in redshift with the others. Similarly, we treat the DES results as independent of all others, as less than $10\%$ of the DES footprint overlaps with BOSS DR12 or eBOSS DR14. The DES sample also consists of very different targets to BOSS and eBOSS, and uses a different methodology (photometric rather than spectroscopic redshifts). However, the BOSS DR12 and eBOSS results are correlated: The eBOSS LRG sample actually contains some of the same galaxies as BOSS DR12, while the eBOSS QSO sample overlaps substantially with the LRGs both on the sky and in redshift. Overall, there are nonzero correlations that should be accounted for between the measurements $(D_{M}/r_{s})^{\mathrm{BOSS}}$, $(Hr_{s})^{\mathrm{BOSS}}$, $(D_{V}/r_{s})^{\mathrm{eBOSS,LRG}}$, and $(D_{V}/r_{s})^{\mathrm{eBOSS,QSO}}$.

To do this in a way that accounts for variation with cosmological parameters, we implement a novel method to compute the cross-correlation coefficients using Fisher matrices, following BAO forecasting techniques \cite{Seo_2007,Font_Ribera_2014}. We sum the Fisher information that each sample contributes to the four overlapping measurements listed above, accounting for redshift and sky overlap. Inverting the full Fisher matrix then gives the correlation coefficients.  We do this separately for every combination of cosmological parameters in the fit, using the number density of objects, matter power spectrum and growth rate of structure to model the BOSS/eBOSS galaxy power spectra and their covariance matrices as a function of redshift. We split the models into smooth and oscillatory components to obtain the derivatives of the BAO feature in the power spectrum with respect to the distance measurements, and fix the galaxy bias and nonlinear damping of the power spectra to their best-fit values as reported in the original works. The Fisher matrix calculation then uses these all as inputs, integrating over angles and scales in the clustering measurements consistent with the range used in the original measurements.

We find a value for the cross-correlation between BOSS DR12 and the eBOSS LRGs comparable to the one reported in Ref.~\cite{Bautista_2018}. The benefits of our technique are that it includes information from all scales included in the measurement and does so self-consistently for each combination of cosmological parameters. More details on the calculation and the computed correlation coefficients can be found in Appendix~\ref{app:BAO}.

We compute the BAO scale and SN Ia likelihoods via an interface to \montepython \textsf{3.3.0} \cite{Audren:2012wb,brinckmann2018montepython}; our novel BAO scale correlation treatment will appear in a future release.  For computing observable predictions, we use routines in \cosmobit and associated interfaces to \alterbbn \textsf{2.2} (\cite{Arbey:2011nf,Arbey:2018zfh}; for BBN yields) and \class \textsf{2.9.3} (\cite{Blas:2011rf}; for solving the background cosmology and Boltzmann equations).

\begin{table}[t]
\centering
\begin{tabular}{l|l|l}
\hline
Sector & Parameter & Range \\
\hline
$\nu$ masses &$m _{\nu_0}$                            & [0, 1.1]\,eV \\
             &$\Delta m^{2}_{21}$                     & $[6,\ 9] \times 10^{-5}$\,eV$^2$ \\
 $\,\,$(NH)  &$\Delta m^{2}_{3l}$                  & $[2.2,\ 2.8] \times 10^{-3}$\,eV$^2$ \\
 $\,\,$(IH)  &$\Delta m^{2}_{3l}$                  & $[-2.8,\ -2.2] \times 10^{-3}$\,eV$^2$ \\
\hline
 \lcdm       &$H_0 $                                  & [50, 80]\,km\,s$^{-1}$\,Mpc$^{-1}$ \\
             &$\Omega_\mathrm{b} h^2$                 & [0.020, 0.024] \\
             &$\Omega_\mathrm{cdm} h^2$               & [0.10, 0.15] \\
             &$\tau_\mathrm{reionization} $           & [0.004, 0.20] \\
             &$\mathrm{ln}\left(10^{10}\,A_s\right) $ & [2.5,\ 3.5] \\
             &$n_s $                                  & [0.90, 1.10] \\
\hline
$N_\mathrm{eff}$ &$r_\nu$                             & [0.75, 1.15] \\
\hline
Nuisance     & SN Ia abs.\ magnitude $M$              & [$-$20, $-$18] \\
             & Neutron lifetime $\tau_\mathrm{n}$     & [876, 883]\,s\\
             & \textit{Planck} likelihood             & 21 parameters varied \\
\hline
\end{tabular}
\caption{Parameters and ranges varied in the main analysis of this article. All parameters are sampled with linear priors. For the nuisance parameters associated with the \textit{Planck} likelihood, we adopt the same prior ranges as done for the Planck baseline analysis \cite{Aghanim:2019ame} and apply the recommended Gaussian priors as additional likelihood contributions.}
\label{tab:priors}
\end{table}

We perform separate fits of the NH and IH, varying $\mnuzero$, $\Delta m_{21}^2$ and $\Delta m_{3l}^2$, the six free parameters of the standard Lambda cold dark matter cosmology (\lcdm; see \cite{cosmobit} for detailed definitions), and $\Delta N_\mathrm{eff}\equiv N_\mathrm{eff} - N_\mathrm{SM}$ with $ N_\mathrm{SM}= 3.045$ \cite{deSalas:2016ztq,Akita:2020szl,Froustey:2020mcq} (\autoref{tab:priors}). For our main analysis, we conservatively adopt a linear prior on $\mnuzero$ between 0 and 1.1\,eV. In Appendix \ref{app:prior}, we show how the limit on $\mnuzero$ strengthens if we change to a logarithmic prior above $\mnuzero=0.0003$\,eV. We adopt linear priors on the \lcdm parameters, as these are sufficiently well constrained by data that their priors are inconsequential.

We assume that $N_\mathrm{eff}$ has the same value during BBN and recombination, varying $\Delta N_\mathrm{eff}$ by scanning linearly over the effective neutrino temperature ratio $r_{\nu,\mathrm{eff}} \equiv T_\nu / T_{\nu,\text{SM}}= (\Delta N_\mathrm{eff}/N_\mathrm{SM} + 1)^\frac14$. This approach enables us to explore the full range of the number of effective relativistic degrees of freedom in the early Universe, corresponding to both positive and negative values of $\Delta N_\mathrm{eff}$. We later explore the impact of restricting our analysis to $\Delta N_\mathrm{eff} \ge 0$, in which case, $\Delta N_\mathrm{eff}$ can be interpreted as the contribution of additional ultrarelativistic (radiation) species and to the pure SM case ($\Delta N_\mathrm{eff} = 0$).

Finally, we vary a total of 23 nuisance parameters representing uncertainties in the SN Ia absolute magnitude, the \textit{Planck} analysis, and the neutron lifetime. These nuisance parameters are constrained, respectively, with the \montepython likelihood for the magnitude of Pantheon supernovae, a likelihood implementation of the \textit{Planck} nuisance priors \cite{Aghanim:2019ame}, and the combination of all ``bottle'' measurements of the lifetime of the neutron $\tau_{n\,,\text{bottle}} = 879.4 \pm 0.6$\,s \cite{PDG20}. We employ the nested sampler \polychord \textsf{1.17.1} \cite{Handley:2015} in fast-slow mode via \scannerbit \cite{scannerbit} in order to oversample \textit{Planck} and SN Ia nuisances. Our fits use 500 live points, 5000 initial samples from the prior, a stopping tolerance of 0.01, $n_\mathrm{repeats}=2n_\mathrm{slow}=22$, a 1:3 fast-slow timing split (leading to approximately 340 likelihood evaluations with different nuisance parameters per combination of the remaining parameters), and default values for all other settings.

\section{Results}

Assuming \lcdm cosmology plus a free $N_\mathrm{eff}$ and normal mass ordering, we find a global 95\% confidence upper bound on the lightest neutrino mass of $\mnuzero < 0.037$\,eV; for inverted ordering, this increases slightly to $\mnuzero < 0.042$\,eV.  In terms of the sum of neutrino masses, this corresponds to $0.058 < \sum m_\nu / \mathrm{eV} < 0.139$ (NH) and $0.098 < \sum m_\nu / \mathrm{eV} < 0.174$ (IH).

\begin{figure}[t]
	\centering
	\includegraphics[width=\columnwidth,clip,trim=5 8 5 0]{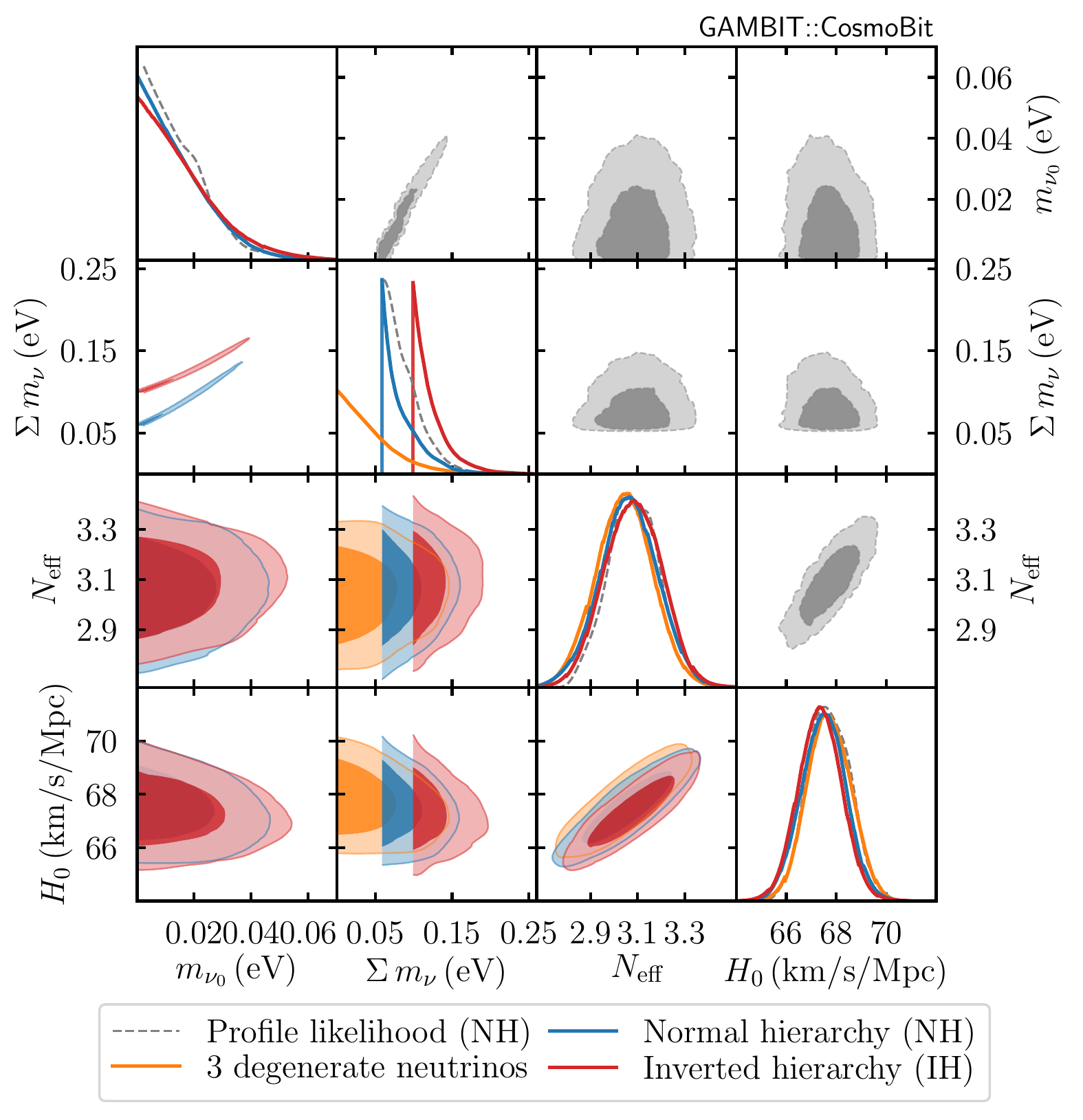}
	\caption{1D and joint 2D posteriors (bottom left) and profile likelihoods (upper right) on the lightest neutrino mass, the sum of neutrino masses, the number of effective neutrino species at the time of CMB formation, and the Hubble parameter, based on the most robust and complete combination to date of CMB, BAO scale, SN Ia, BBN and neutrino oscillation data. Posteriors are shown for normal and inverted neutrino hierarchies and are compared with the often-seen (but unphysical) scenario of three degenerate-mass neutrinos. Profile likelihoods are for the normal hierarchy only and assume the best-fit values from the corresponding posterior scan for the 21 \textit{Planck} nuisance parameters. Shading indicates 68\% and 95\% credible/confidence regions.
		\label{fig1}}
\end{figure}

The lower-left triangle of \autoref{fig1} compares the one- and two-dimensional marginalized posterior distributions for the normal and inverted hierarchies. Here we show both the lightest neutrino mass and the sum of neutrino masses, as well as their correlations with $N_\mathrm{eff}$ and $H_0$. We also show the result for the canonical scenario considered in most previous analyses, such as those by both \textit{Planck} \cite{Aghanim:2018eyx} and eBOSS \cite{Alam:2020sor}, where a single parameter specifies a common degenerate mass for all three neutrinos. In all cases, the maximum posterior probability density is achieved for a massless lightest neutrino, reflecting the fact that there remains no positive cosmological hint for neutrino mass to date. The degenerate-mass assumption would lead one to erroneously infer that $\sum m_\nu / \mathrm{eV} < 0.115$ at 95\% confidence.  This result is plainly biased toward lower values due to the fact that the majority of the probability distribution lies within the unphysical region excluded by neutrino oscillation experiments. Using a physically realistic mass model shifts the 95\% interval for $H_0$ from $67.7\pm 1.7$\,km\,s$^{-1}$\,Mpc$^{-1}$ to $67.5\pm 1.8$\,km\,s$^{-1}$\,Mpc$^{-1}$ (NH) or $67.4\pm 1.7$\,km\,s$^{-1}$\,Mpc$^{-1}$ (IH), and the 95\% interval for $N_\mathrm{eff}$ from $3.04\pm 0.24$ to $3.06\pm 0.24$ (NH) or $3.08\pm 0.24$ (IH).

For comparison, in the upper-right triangle of \autoref{fig1}, we also show prior-independent profile likelihoods for the NH obtained from the differential evolution sampler \textsf{Diver} \cite{scannerbit} with a population of $10^4$, a convergence threshold of $10^{-4}$, and the \textit{Planck} nuisance parameters fixed to their best-fit values from the NH \polychord fit (as \textsf{Diver} has no fast-slow feature). The results match the posteriors reasonably closely, but give slightly stronger implied constraints at 95\% confidence: $\mnuzero < 0.033$\,eV and $0.058 < \sum m_\nu / \mathrm{eV} < 0.127$.\footnote{The slightly higher profile likelihood than posterior for much of the allowed range of $\sum m_\nu$ should be understood in the context of frequentist confidence levels deriving from isolikelihood contours, rather than integrated posterior probabilities as in the case of Bayesian credible regions.} These findings confirm the robustness of our main (Bayesian) results. We have also calculated Bayesian evidences and find that the IH is disfavored relative to the NH with Bayes factors between $\log B = 5.6$ and $\log B = 7.0$ (depending on the treatment of $N_\text{eff}$), a result driven mostly by the neutrino oscillation likelihoods~\cite{deSalas:2020pgw}.

\begin{figure}[t]
\centering
\includegraphics[width=0.95\columnwidth]{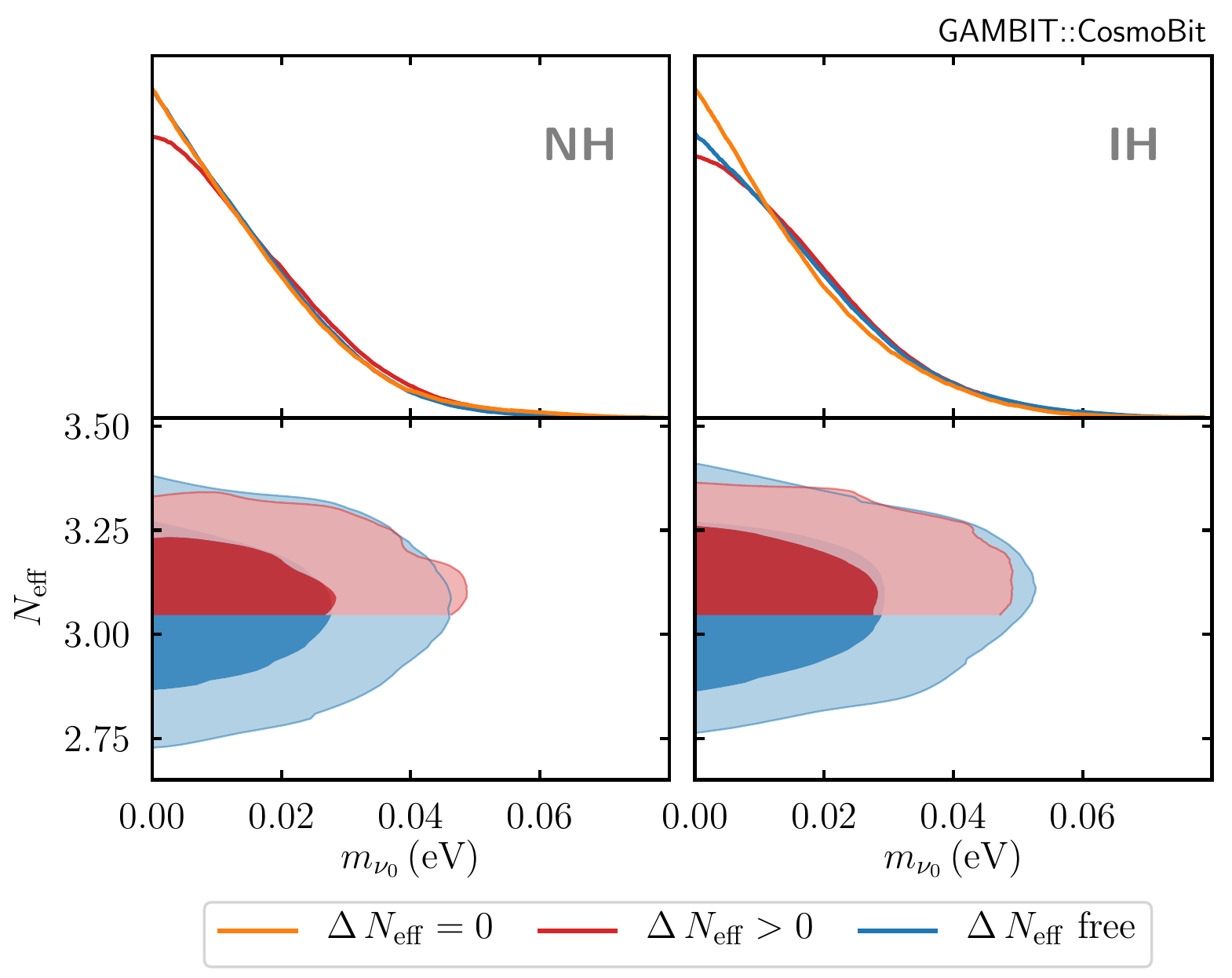}
\caption{Comparison of posterior probabilities for the mass of the lightest neutrino under three different assumptions: $\Delta N_\mathrm{eff} = 0$, $\Delta N_\mathrm{eff} \ge 0$ corresponding to dark radiation, or $\Delta N_\mathrm{eff}$ free to take on positive or negative values, corresponding to a modified effective neutrino temperature.}
\label{fig2}
\end{figure}

In \autoref{fig2}, we examine the impacts of different physical sources of $N_\mathrm{eff}$: (a) changes in the neutrino temperature, where $\Delta N_\mathrm{eff}$ is allowed to be positive or negative, as in our benchmark analyses, (b) dark radiation, where $\Delta N_\mathrm{eff}\ge0$, and (c) the pure SM case, where $\Delta N_\mathrm{eff}=0$. The resulting posteriors only change very slightly, corresponding to shifts of the order of 0.002--0.003\,eV in the 95\% limit on $\mnuzero$.  Our final results can therefore be considered rather robust to assumptions about $\Delta N_\mathrm{eff}$. \autoref{fig2} suggests that at the 68\% confidence level, allowing positive $\Delta N_\mathrm{eff}$ may weaken the limits slightly compared to $\Delta N_\mathrm{eff} = 0$, with the effect offset to some extent by also allowing $\Delta N_\mathrm{eff} < 0$.  Any such effect is however small enough that it is difficult to distinguish from sampling noise. Notably, bounds on $\mnuzero$ and $\sum m_\nu$ can be weakened substantially in cosmologies featuring \textit{both} dark radiation and a modified neutrino temperature \cite{cosmobit}, neutrino self-interactions, or exotic dark energy, even to the level where a direct measurement of the neutrino mass may be within reach of the KATRIN experiment~\cite{Aker:2019uuj}.

\section{Discussion}

Our analysis provides a more precise and robust limit on the sum of neutrino masses than either those of \textit{Planck} \cite{Aghanim:2018eyx} or eBOSS \cite{Alam:2020sor}, mainly due to our use of physical neutrino mass models rather than the assumption of degenerate masses. Comparing results for the unphysical degenerate-mass model however provides an indication of the constraining power of the cosmological data used in each case. Our limit ($\sum m_\nu / \mathrm{eV} < 0.115$) is very close to the most similar combination in Ref.\ \cite{Aghanim:2018eyx} ($\sum m_\nu / \mathrm{eV} < 0.12$, \textit{Planck} TT,TE,EE+lowE+lensing+BAO, with varying $N_{\rm{eff}}$), indicating that within $\Lambda$CDM+$N_\text{eff}$ the eBOSS DR14 and DES BAO and Pantheon SN Ia measurements only add limited additional information.

The most similar eBOSS limit ($\sum m_\nu / \mathrm{eV} < 0.099$) is stronger. Reference \cite{Alam:2020sor} includes slightly more up-to-date eBOSS BAO scale measurements compared to our analysis, but the improved sensitivity is mainly driven by their inclusion of redshift-space distortions (RSD).  We do not include RSD measurements, as to date they have been based on templates for the matter power spectrum that assume a particular neutrino mass, and the fits then neglect the scale dependence in the growth rate of structure. As such we believe they cannot robustly be used to constrain neutrino masses. Reference \cite{Alam:2020sor} also includes new (DR16) Ly$\alpha$ constraints on the BAO scale. We checked that including DR14 Ly$\alpha$ measurements has no impact on our limit. Given this result, the fact that the redshifts probed by Ly$\alpha$ data are intermediate between those of the CMB and other BAO measurements, that Ly$\alpha$ BAO are slightly discrepant with other BAO, and that the Ly$\alpha$ BAO results require more precise control over observational and astrophysical systematics than galaxy BAO \cite{Blomqvist2019}, we argue that excluding the DR14 Ly$\alpha$ result gives a more robust limit on neutrino masses at no decrease in the statistical constraining power.

The limits that we present here on the mass of the lightest neutrino are almost 60\% stronger than those derived from a combined fit to both mass hierarchies in a recent similar analysis ($\mnuzero < 0.086$\,eV \cite{Loureiro:2018pdz}), and 14\% stronger than those appearing in a similar contemporaneous analysis \cite{deSalas:2020pgw}. At the limiting value of $\mnuzero$, this brings the absolute scale of neutrino masses down to a level comparable to the larger of the two mass splittings ($m_3 - m_1=0.025$\,eV in the NH, $m_2 - m_3=0.023$\,eV in the IH). We use the same BBN and SN~Ia data as Ref.\ \cite{Loureiro:2018pdz} but improved neutrino and CMB data: results from \nufit \textsf{4.1} rather than \textsf{2.1} for neutrino experiments (leading to roughly 20\% stronger bounds), and a CMB likelihood based on 2018 rather than 2015 \textit{Planck} data (leading to roughly 30\% stronger bounds).  Compared to \cite{deSalas:2020pgw}, we add BBN constraints, DES, and eBOSS BAO scale data and marginalize over $N_\mathrm{eff}$. We also propagate the primordial helium abundance fully and incorporate the uncertainty on the lifetime of the neutron (see also discussion in Ref.\ \cite{cosmobit}). From galaxy surveys, we and Ref.\ \cite{deSalas:2020pgw} rely exclusively on scale measurements, as a correct statistical combination of scale data can provide a strong limit on neutrino masses that is also very robust. Reference \cite{Loureiro:2018pdz} instead used an angular clustering reanalysis of the BOSS DR12 data.  Their approach used the full shape of the clustering more self-consistently than template-based RSD measurements but still required several uncertainties such as the galaxy bias, redshift error dispersions, and spectroscopic redshift errors to be modeled, and data cuts to be made, adding 28 more nuisance parameters. In principle, nonlinear scales should provide the most constraining power on neutrino mass parameters but difficulty in the modeling limits the accessibility of this information \cite{Font-Ribera:2013wce}. With current analysis techniques, BAO scale measurements still add more constraining power for neutrino masses than data encoding the full shape of the galaxy power spectrum---although this is not expected to remain true for much longer \cite{Ivanov:2019hqk}.

\section{Summary}

We have presented a comprehensive combined analysis of recent neutrino oscillation, CMB, SN Ia, BBN, and BAO scale data deriving the most accurate and precise limits to date on the mass of the lightest neutrino and the sum of neutrino masses.  Assuming normal mass ordering and standard cosmology plus $\Delta N_\mathrm{eff}\ne0$, we have found $\mnuzero < 0.037$\,eV and $0.058 < \sum m_\nu / \mathrm{eV} < 0.139$.  With inverted ordering, $\mnuzero < 0.042$\,eV and $0.098 < \sum m_\nu / \mathrm{eV} < 0.174$.  These results should serve as a benchmark in the coming years, as neutrino cosmology continues its inexorable progress toward a measurement of the absolute neutrino mass scale.

All input files and parameter samples produced for this article can be found on \textsf{Zenodo}
\cite{CosmoBit_numass_zenodo}.

\section*{Acknowledgements}

We thank the \gambit Community, Alex Arbey, Thejs Brinckmann, Tamara Davis, Martina Gerbino, Deanna Hooper, Julien Lesgourgues, Vivian Poulin, Nils Sch\"{o}neberg, Jes\'{u}s Torrado and Sunny Vagnozzi for helpful discussions, PRACE for access to Joliot-Curie at CEA, RWTH Aachen University for access to JARA under Project No. jara0184 and the University of Cambridge for access to CSD3 resources.  P.~St. and F.~K. acknowledge funding from Deutsche Forschungsgemeinschaft Grant No.~KA~4662/1-1, C.~B., T.~E.~G., M.~J.~W., P.~Sc. and C.~H. from Australian Research Cuncil Grants No.~DP180102209, No.~FL180100168, No.~CE200100008, and No.~FT190100814, J.~J.~R. from Swedish Research Council Contract No.~638-2013-8993, W.~H. from a Gonville \& Caius Research Fellowship and the George Southgate visiting fellowship, and A.~C.~V. from the Arthur B. McDonald Canadian Astroparticle Physics Research Institute, Canada Foundation for Innovation and Ontario Ministry of Economic Development, Job Creation and Trade (MEDJCT). Research at Perimeter Institute is supported by the Government of Canada through the Department of Innovation, Science, and Economic Development, and by the Province of Ontario through MEDJCT. This article made use of \textsf{matplotlib}~\cite{Hunter:2007}, \textsf{GetDist}~\cite{Lewis:2019xzd}, and \textsf{pippi}~\cite{pippi}.

\appendix

\section{Details of BAO scale correlation coefficients}
\label{app:BAO}

Here we provide more detail on our novel method to account for the correlations between overlapping BAO experiments in a cosmology-dependent way. We acknowledge that the use of cosmology-dependent Gaussian covariance matrices can lead to an overestimation of the Fisher information \citep{Carron_2013}; however, in our method we are only computing the cross-correlation coefficients, which are then scaled by the original measurement errors. We do not modify these original errors and so are not underestimating the uncertainty. More so, we argue that including cosmology dependence in the correlation coefficients is the correct thing to do, as the overlap between BAO measurements is mainly related to the (effective) cosmological volume shared by the experiments, which is clearly a function of the cosmological model.

We begin by expressing the joint covariance matrix of BAO measurements from the overlapping BOSS DR12 and eBOSS results $\boldsymbol{\mathsf{C}}$ in terms of the inverse of the Fisher matrix $\boldsymbol{\mathsf{F}}$,
\begin{equation}
\boldsymbol{\mathsf{F}}^{-1} = \boldsymbol{\mathsf{C}} = \boldsymbol{\mathsf{E}}
\begin{pmatrix}
1 & \rho^\mathrm{BOSS}_{D_{M}/r_{s},Hr_{s}} & c_{0} & 0 \\
\rho^\mathrm{BOSS}_{D_{M}/r_{s},Hr_{s}} & 1 & c_{1} & 0 \\
c_{0} & c_{1} & 1 & c_{2} \\
0 & 0 & c_{2} & 1 \\
\end{pmatrix} \boldsymbol{\mathsf{E}}.
\end{equation}
Here, $c_0$ is the correlation coefficient between measurements $(D_{M}/r_{s})^{\mathrm{BOSS}}$ and $(D_{V}/r_{s})^{\mathrm{eBOSS,LRG}}$, $c_1$ is the correlation between $(Hr_{s})^{\mathrm{BOSS}}$ and $(D_{V}/r_{s})^{\mathrm{eBOSS,LRG}}$, and $c_2$ is the correlation between $(D_{V}/r_{s})^{\mathrm{eBOSS,LRG}}$ and $(D_{V}/r_{s})^{\mathrm{eBOSS,QSO}}$. The cross-correlation $\rho^{BOSS}_{D_{M}/r_{s},Hr_{s}}$ between the BOSS DR12 measurements of $D_{M}/r_{s}$ and $Hr_{s}$ is provided as part of the BOSS DR12 results in \montepython and is not replaced in this analysis, or made dependent upon cosmological parameters. The matrix $\boldsymbol{\mathsf{E}}=\mathrm{diag}(\sigma^\mathrm{BOSS}_{D_{M}/r_{s}}, \sigma^\mathrm{BOSS}_{Hr_{s}}, \sigma^\mathrm{eBOSS,LRG}_{D_{V}/r_{s}}, \sigma^\mathrm{eBOSS,QSO}_{D_{V}/r_{s}})$ contains the experimental uncertainties from the BAO measurements, which are also included in \montepython. Hence the only unknowns are $c_{i}$, which we calculate for each set of cosmological parameters by inverting the full Fisher matrix. In practice, we compute the Fisher matrix for the BAO scale parameters
\begin{equation}
\alpha=\frac{D_{V}r^{\mathrm{fid}}_{s}}{D^{\mathrm{fid}}_{V}r_{s}}, \quad \alpha_{\perp}=\frac{D_{M}r^{\mathrm{fid}}_{s}}{D^{\mathrm{fid}}_{M}r_{s}}, \quad \alpha_{||}=\frac{H^{\mathrm{fid}}r^{\mathrm{fid}}_{s}}{Hr_{s}},
\end{equation}
where ``fid'' corresponds to the fiducial cosmology used to make the original clustering measurements. It is trivial to convert the correlation coefficients to those for the distance scales that we actually fit in \montepython.

We do this based on \cite{Howlett_2016}, writing the Fisher matrix elements for each parameter of interest as a sum over the information from the different redshift bins and overlapping/nonoverlapping sky areas for each survey. For example, consider the redshift range $0.60 < z < 0.75$. For each redshift bin in this range, we have information contributing to three parameters: $\alpha^{\mathrm{BOSS}}_{\perp}$, $\alpha_{||}^{\mathrm{BOSS}}$ and $\alpha^{\mathrm{eBOSS,LRG}}$. We first compute the $2 \times 2$ Fisher ``submatrix'' for $\alpha^{\mathrm{BOSS}}_{\perp}$, $\alpha_{||}^{\mathrm{BOSS}}$ from the nonoverlapping sky area $\Omega^{\mathrm{BOSS}}-\Omega^{\mathrm{eBOSS,LRG}}$ and add this to the full matrix, then add on the $3 \times 3$ Fisher submatrix for all three parameters from the sky area $\Omega^{\mathrm{eBOSS,LRG}}$. 

The Fisher matrix element for parameters $\lambda_{i}$ and $\lambda_{j}$ measured from survey $A$ within a redshift bin is \cite{Tegmark_1997}
\begin{equation}
F^{A}_{ij} = \int_{0}^{k_{\mathrm{max}}} k^{2} dk \int_{0}^{1} d\mu \frac{\partial P^{A}_{g}(k,\mu)}{\partial \lambda_{i}} C^{-1}_{P^{A}_{g}}(k,\mu) \frac{\partial P^{A}_{g}(k,\mu)}{\partial \lambda_{j}},
\label{eq:fisher}
\end{equation}
where we fix $k_{\mathrm{max}}=0.3h\,\mathrm{Mpc^{-1}}$. $C_{P^{A}_{g}}(k,\mu)$ is the covariance matrix of the galaxy power spectrum $P^{A}_{g}(k,\mu)$,
\begin{equation}
C_{P^{A}_{g}}(k,\mu) =  \frac{4\pi^{2}}{V^{A}}\biggl[P^{A}_{g}(k,\mu)+\frac{V^{A}}{N^{A}}\biggl]^{2},
\end{equation}
where $N^{A}$ is the number of galaxies in the redshift bin, and $V^{A}$ is the cosmological volume contained within the redshift bin and sky area $\Omega^{A}$. The first term in the covariance matrix models cosmic variance; the second is shot noise from the finite number of galaxies. We model the galaxy power spectrum as a function of the matter power spectrum $P_{m}(k)$ and potentially scale-dependent growth rate of structure $f(k)$, splitting it into two components using the smoothed matter power spectrum $P_{sm}(k)$,
\begin{multline}
P^{A}_{g}(k,\mu) = (b^{A}+f(k)\mu^{2})^{2}P_{sm}(k) \\
\times \biggl[1+\biggl(\frac{P_{m}(k)}{P_{sm}(k)}-1\biggl)e^{-\frac{1}{2}k^{2}[\mu^{2}\Sigma^{A}_{nl,||} + (1 - \mu^{2})\Sigma^{A}_{nl,\perp}]}\biggl].
\end{multline}
Here $b^{A}$ represents the linear galaxy bias for the survey, and $\Sigma^{A}_{nl,||}$ and $\Sigma^{A}_{nl,\perp}$ account for nonlinear damping of the BAO feature. The values for these are fixed to the best-fit values from the BOSS and eBOSS analyses. We compute $P_{sm}(k)$ using the method of Ref.\ \cite{Hinton_2016}.

\begin{figure}[t]
\centering
\includegraphics[width=.975\columnwidth]{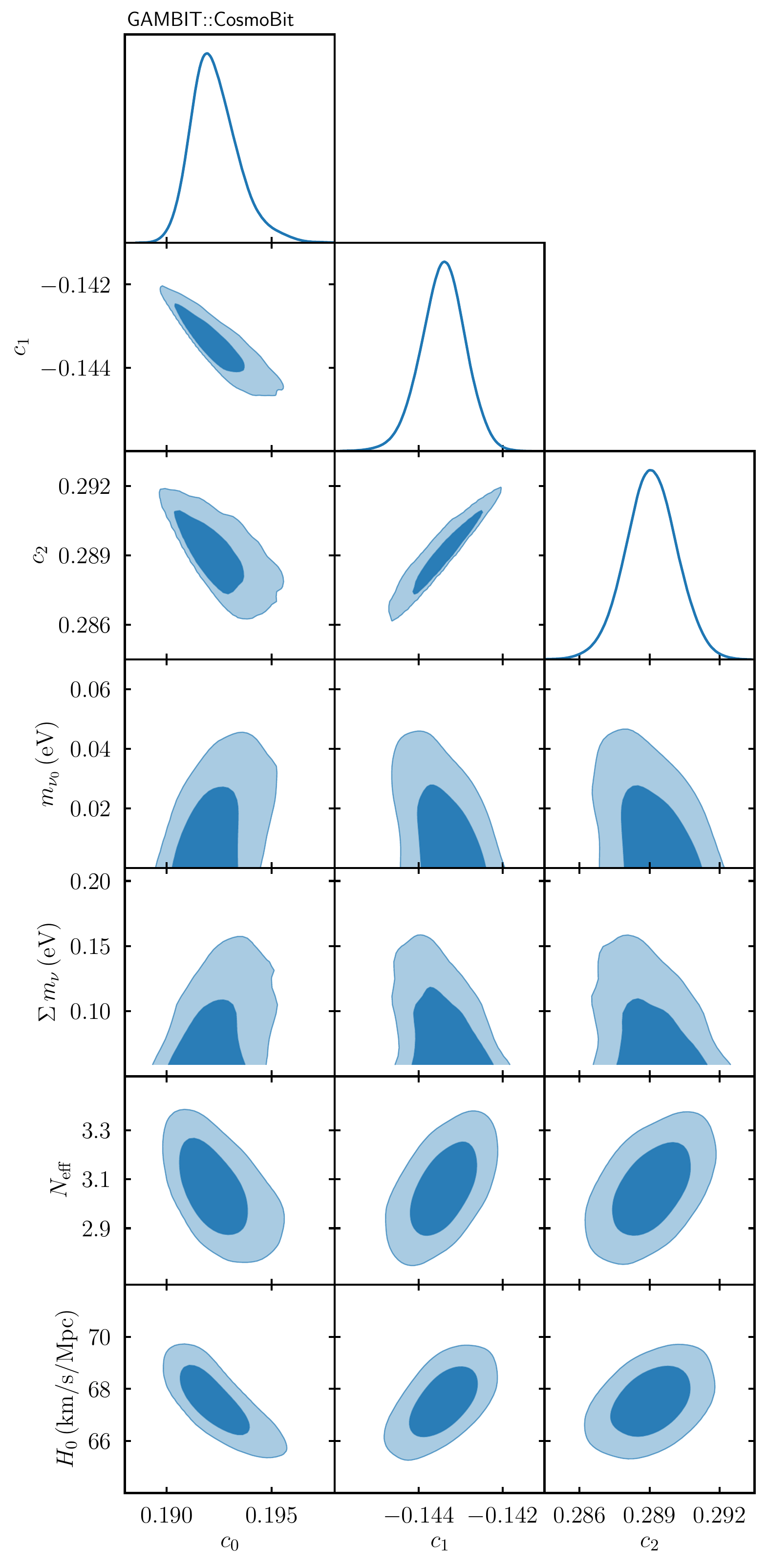}
\caption{1D and joint 2D posteriors for the correlation coefficients $c_0$ [$(D_{M}/r_{s})^{\mathrm{BOSS}}$ and $(D_{V}/r_{s})^{\mathrm{eBOSS,LRG}}$], $c_1$ [$(Hr_{s})^{\mathrm{BOSS}}$ and $(D_{V}/r_{s})^{\mathrm{eBOSS,LRG}}$], and $c_2$ [$(D_{V}/r_{s})^{\mathrm{eBOSS,LRG}}$ and $(D_{V}/r_{s})^{\mathrm{eBOSS,QSO}}$].  Also shown are the variations of the correlation coefficients with neutrino mass parameters $N_\mathrm{eff}$ and the Hubble parameter $H_0$.
\label{fig3}}
\end{figure}

The split into smooth and nonsmooth components of the matter power spectrum ensures that we are only including information from the BAO scale, and not the broadband shape of the power spectrum or redshift-space distortions. As such, the derivatives with respect to $\lambda_{i,j}$ in Eq.~(\ref{eq:fisher}) are computed only on the $P_{m}(k)/P_{sm}(k)$ component of $P^{A}_{g}(k,\mu)$. We do this by finite differencing $P_{m}(k')/P_{sm}(k')$ evaluated at $k'=k/\alpha$ or
\begin{equation}
k' = \frac{k}{\alpha_{\perp}}\biggl[1 + \mu^{2}\biggl(\frac{\alpha^{2}_{\perp}}{\alpha^{2}_{||}}-1\biggl)\biggl]^{1/2}.
\end{equation}
Overall, the calculation of the Fisher matrix for a particular redshift bin and survey matches that commonly used in the literature \cite{Seo_2007,Font_Ribera_2014} and our model power spectrum is representative of how BAO constraints are actually extracted from the data \cite{Alam:2016hwk,Ata:2017dya}.

In \autoref{fig3}, we show the distribution of the correlation coefficients in our BAO scale joint likelihood from our main NH fit.  The variation of the correlation coefficients with cosmological parameters is small but perceptible, with $c_1$ and $c_2$ both increasing along with $H_0$ and $N_\mathrm{eff}$, and $c_0$ decreasing with larger $H_0$ and $N_\mathrm{eff}$. The trends are weaker but in the opposite direction for the neutrino mass parameters.  In general $c_1$ and $c_2$ are strongly correlated with each other and anticorrelated with $c_0$. Our method can be easily extended to include other datasets, and we expect larger variation in the values of the coefficients for models including nonzero curvature or non-cosmological-constant models of dark energy.

\section{Impacts of priors on the mass of the lightest neutrino}
\label{app:prior}

\begin{figure}[b]
	\centering
	\includegraphics[width=\columnwidth]{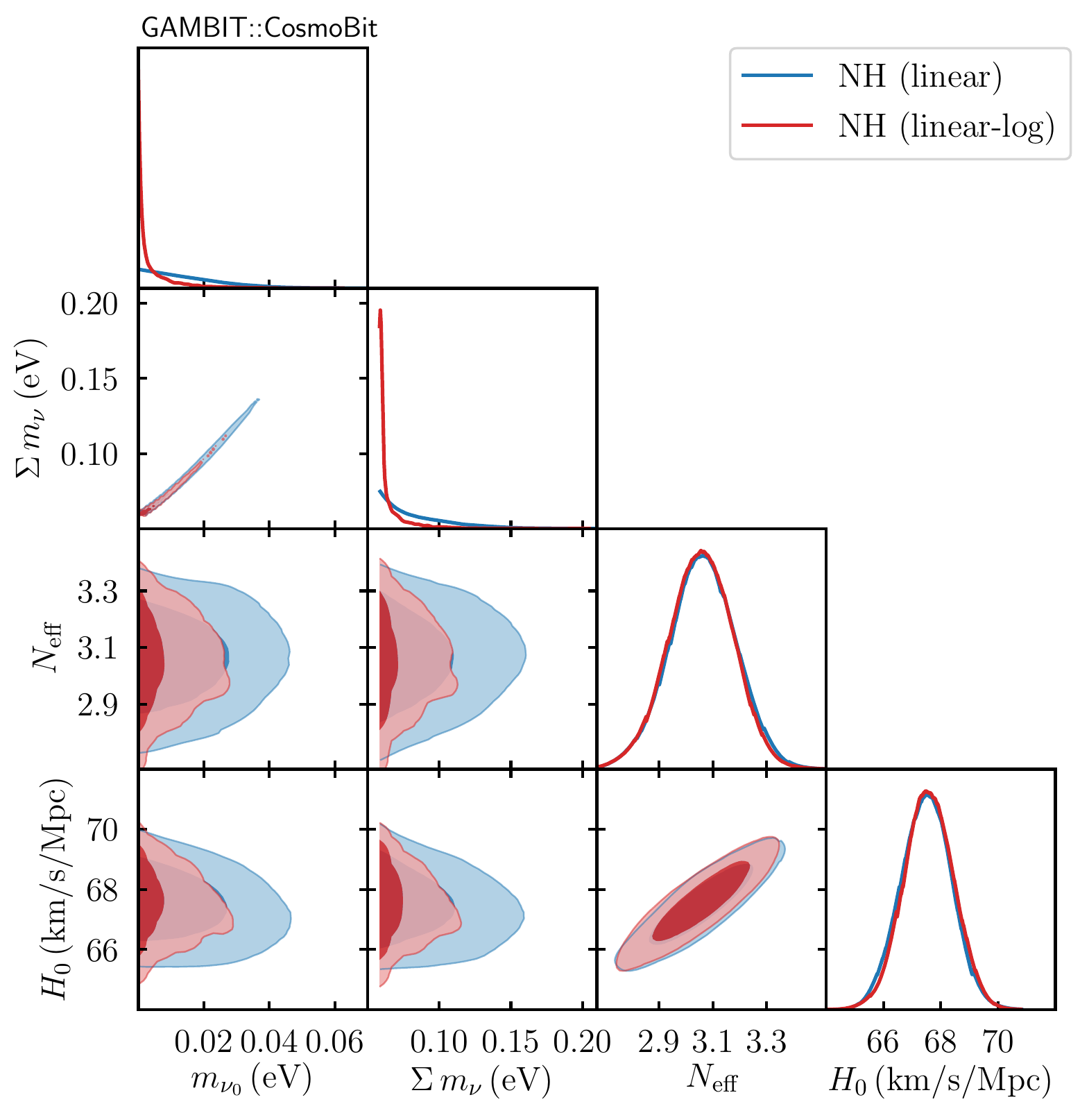}
	\caption{1D and joint 2D posteriors on the neutrino mass and relevant cosmological parameters, assuming normal ordering. Here we reproduce the results from \autoref{fig1} based on a linear prior for the mass of the lightest neutrino and compare them to the results when employing a hybrid prior, linear below $\mnuzero=0.0003$\,eV and logarithmic above.}
	\label{fig4}
\end{figure}

In the main body of this article, we presented only results based on a linear prior for the lightest neutrino mass $\mnuzero$.  This is a conservative choice, as it is an essentially uninformative prior (but see Ref.\ \cite{Heavens:2018adv} for an objective Bayesian construction of the most uninformative prior for this problem). Although more informative, a logarithmic prior is arguably more physically justified: Given that we have no information suggesting any preferred scale for $\mnuzero$ (or, in fact any preference for $\mnuzero>0$ at all), one might argue that any scale for the mass is as likely as another.  In \autoref{fig4} we compare our result for the NH with the result if we instead adopt a logarithmic prior for $\mnuzero$ above $0.0003$\,eV (and retain the linear prior below this value, enforcing continuity of the prior across the transition).  Masses below 0.0003\,eV are indistinguishable from the massless case in the outputs of \class, so there is little point in oversampling this region. As can be seen from \autoref{fig4}, this hybrid linear-logarithmic prior indeed results in a much stronger preference for very small neutrino masses, giving $\mnuzero < 0.020$\,eV and $0.058 < \sum m_\nu / \mathrm{eV} < 0.100$ at 95\% confidence.  The effect is similar in the IH, producing $\mnuzero < 0.024$\,eV and $0.098 < \sum m_\nu / \mathrm{eV} < 0.136$ at 95\% confidence.

\vspace{4cm}

\bibliography{R2}

\begin{thebibliography}{79}%
\makeatletter
\providecommand \@ifxundefined [1]{%
 \@ifx{#1\undefined}
}%
\providecommand \@ifnum [1]{%
 \ifnum #1\expandafter \@firstoftwo
 \else \expandafter \@secondoftwo
 \fi
}%
\providecommand \@ifx [1]{%
 \ifx #1\expandafter \@firstoftwo
 \else \expandafter \@secondoftwo
 \fi
}%
\providecommand \natexlab [1]{#1}%
\providecommand \enquote  [1]{``#1''}%
\providecommand \bibnamefont  [1]{#1}%
\providecommand \bibfnamefont [1]{#1}%
\providecommand \citenamefont [1]{#1}%
\providecommand \href@noop [0]{\@secondoftwo}%
\providecommand \href [0]{\begingroup \@sanitize@url \@href}%
\providecommand \@href[1]{\@@startlink{#1}\@@href}%
\providecommand \@@href[1]{\endgroup#1\@@endlink}%
\providecommand \@sanitize@url [0]{\catcode `\\12\catcode `\$12\catcode
  `\&12\catcode `\#12\catcode `\^12\catcode `\_12\catcode `\%12\relax}%
\providecommand \@@startlink[1]{}%
\providecommand \@@endlink[0]{}%
\providecommand \url  [0]{\begingroup\@sanitize@url \@url }%
\providecommand \@url [1]{\endgroup\@href {#1}{\urlprefix }}%
\providecommand \urlprefix  [0]{URL }%
\providecommand \Eprint [0]{\href }%
\providecommand \doibase [0]{https://doi.org/}%
\providecommand \selectlanguage [0]{\@gobble}%
\providecommand \bibinfo  [0]{\@secondoftwo}%
\providecommand \bibfield  [0]{\@secondoftwo}%
\providecommand \translation [1]{[#1]}%
\providecommand \BibitemOpen [0]{}%
\providecommand \bibitemStop [0]{}%
\providecommand \bibitemNoStop [0]{.\EOS\space}%
\providecommand \EOS [0]{\spacefactor3000\relax}%
\providecommand \BibitemShut  [1]{\csname bibitem#1\endcsname}%
\let\auto@bib@innerbib\@empty
\bibitem [{\citenamefont {Gando}\ \emph {et~al.}(2013)\citenamefont {Gando}
  \emph {et~al.}}]{Gando:2013nba}%
  \BibitemOpen
  \bibfield  {author} {\bibinfo {author} {\bibfnamefont {A.}~\bibnamefont
  {Gando}} \emph {et~al.} (\bibinfo {collaboration} {KamLAND}),\ }\href
  {https://doi.org/10.1103/PhysRevD.88.033001} {\bibfield  {journal} {\bibinfo
  {journal} {\prd}\ }\textbf {\bibinfo {volume} {88}},\ \bibinfo {pages}
  {033001} (\bibinfo {year} {2013})},\ \Eprint
  {https://arxiv.org/abs/1303.4667} {arXiv:1303.4667} \BibitemShut {NoStop}%
\bibitem [{\citenamefont {An}\ \emph {et~al.}(2017)\citenamefont {An} \emph
  {et~al.}}]{An:2016srz}%
  \BibitemOpen
  \bibfield  {author} {\bibinfo {author} {\bibfnamefont {F.~P.}\ \bibnamefont
  {An}} \emph {et~al.} (\bibinfo {collaboration} {Daya Bay}),\ }\href
  {https://doi.org/10.1088/1674-1137/41/1/013002} {\bibfield  {journal}
  {\bibinfo  {journal} {Chin. Phys.}\ }\textbf {\bibinfo {volume} {C41}},\
  \bibinfo {pages} {013002} (\bibinfo {year} {2017})},\ \Eprint
  {https://arxiv.org/abs/1607.05378} {arXiv:1607.05378} \BibitemShut {NoStop}%
\bibitem [{\citenamefont {Adey}\ \emph {et~al.}(2018)\citenamefont {Adey} \emph
  {et~al.}}]{Adey:2018zwh}%
  \BibitemOpen
  \bibfield  {author} {\bibinfo {author} {\bibfnamefont {D.}~\bibnamefont
  {Adey}} \emph {et~al.} (\bibinfo {collaboration} {Daya Bay}),\ }\href
  {https://doi.org/10.1103/PhysRevLett.121.241805} {\bibfield  {journal}
  {\bibinfo  {journal} {\prl}\ }\textbf {\bibinfo {volume} {121}},\ \bibinfo
  {pages} {241805} (\bibinfo {year} {2018})},\ \Eprint
  {https://arxiv.org/abs/1809.02261} {arXiv:1809.02261} \BibitemShut {NoStop}%
\bibitem [{\citenamefont {Serra}()}]{dchooz}%
  \BibitemOpen
  \bibfield  {author} {\bibinfo {author} {\bibfnamefont {A.~C.}\ \bibnamefont
  {Serra}},\ }\href@noop {} {\bibinfo {title} {{Double Chooz Improved
  Multi-Detector Measurements. \textit{CERN EP colloquium}, CERN, Switzerland,
  September 20, 2016}}}\BibitemShut {NoStop}%
\bibitem [{\citenamefont {Bak}\ \emph {et~al.}(2018)\citenamefont {Bak} \emph
  {et~al.}}]{Bak:2018ydk}%
  \BibitemOpen
  \bibfield  {author} {\bibinfo {author} {\bibfnamefont {G.}~\bibnamefont
  {Bak}} \emph {et~al.} (\bibinfo {collaboration} {RENO}),\ }\href
  {https://doi.org/10.1103/PhysRevLett.121.201801} {\bibfield  {journal}
  {\bibinfo  {journal} {\prl}\ }\textbf {\bibinfo {volume} {121}},\ \bibinfo
  {pages} {201801} (\bibinfo {year} {2018})},\ \Eprint
  {https://arxiv.org/abs/1806.00248} {arXiv:1806.00248} \BibitemShut {NoStop}%
\bibitem [{\citenamefont {Adamson}\ \emph
  {et~al.}(2013{\natexlab{a}})\citenamefont {Adamson} \emph
  {et~al.}}]{Adamson:2013whj}%
  \BibitemOpen
  \bibfield  {author} {\bibinfo {author} {\bibfnamefont {P.}~\bibnamefont
  {Adamson}} \emph {et~al.} (\bibinfo {collaboration} {MINOS}),\ }\href
  {https://doi.org/10.1103/PhysRevLett.110.251801} {\bibfield  {journal}
  {\bibinfo  {journal} {\prl}\ }\textbf {\bibinfo {volume} {110}},\ \bibinfo
  {pages} {251801} (\bibinfo {year} {2013}{\natexlab{a}})},\ \Eprint
  {https://arxiv.org/abs/1304.6335} {arXiv:1304.6335} \BibitemShut {NoStop}%
\bibitem [{\citenamefont {Adamson}\ \emph
  {et~al.}(2013{\natexlab{b}})\citenamefont {Adamson} \emph
  {et~al.}}]{Adamson:2013ue}%
  \BibitemOpen
  \bibfield  {author} {\bibinfo {author} {\bibfnamefont {P.}~\bibnamefont
  {Adamson}} \emph {et~al.} (\bibinfo {collaboration} {MINOS}),\ }\href
  {https://doi.org/10.1103/PhysRevLett.110.171801} {\bibfield  {journal}
  {\bibinfo  {journal} {\prl}\ }\textbf {\bibinfo {volume} {110}},\ \bibinfo
  {pages} {171801} (\bibinfo {year} {2013}{\natexlab{b}})},\ \Eprint
  {https://arxiv.org/abs/1301.4581} {arXiv:1301.4581} \BibitemShut {NoStop}%
\bibitem [{\citenamefont {Izmaylov}()}]{t2k}%
  \BibitemOpen
  \bibfield  {author} {\bibinfo {author} {\bibfnamefont {A.}~\bibnamefont
  {Izmaylov}},\ }\href@noop {} {\bibinfo {title} {{T2K Neutrino Experiment:
  Recent Results and Plans. \textit{Flavour Physics Conference}, Quy Nhon,
  Vietnam, August 13-19, 2017}}}\BibitemShut {NoStop}%
\bibitem [{\citenamefont {Friend}()}]{t2k2}%
  \BibitemOpen
  \bibfield  {author} {\bibinfo {author} {\bibfnamefont {M.}~\bibnamefont
  {Friend}},\ }\href@noop {} {\bibinfo {title} {{Updated Results from the T2K
  Experiment with $3.13 \times 10^{21}$ Protons on Target. \textit{KEK
  seminar}, January 10, 2019}}}\BibitemShut {NoStop}%
\bibitem [{\citenamefont {Sanchez}(2018)}]{sanchez_mayly_2018_1286758}%
  \BibitemOpen
  \bibfield  {author} {\bibinfo {author} {\bibfnamefont {M.}~\bibnamefont
  {Sanchez}},\ }\href {https://doi.org/10.5281/zenodo.1286758} {\bibinfo
  {title} {Nova results and prospects}} (\bibinfo {year} {2018}),\ \bibinfo
  {note} {\textit{Neutrino 2018}, Heidelberg, June 4, 2018.}\BibitemShut
  {Stop}%
\bibitem [{\citenamefont {Acero}\ \emph {et~al.}(2019)\citenamefont {Acero}
  \emph {et~al.}}]{Acero:2019ksn}%
  \BibitemOpen
  \bibfield  {author} {\bibinfo {author} {\bibfnamefont {M.}~\bibnamefont
  {Acero}} \emph {et~al.} (\bibinfo {collaboration} {NOvA}),\ }\href
  {https://doi.org/10.1103/PhysRevLett.123.151803} {\bibfield  {journal}
  {\bibinfo  {journal} {\prl}\ }\textbf {\bibinfo {volume} {123}},\ \bibinfo
  {pages} {151803} (\bibinfo {year} {2019})},\ \Eprint
  {https://arxiv.org/abs/1906.04907} {arXiv:1906.04907} \BibitemShut {NoStop}%
\bibitem [{\citenamefont {Cleveland}\ \emph {et~al.}(1998)\citenamefont
  {Cleveland}, \citenamefont {Daily}, \citenamefont {Davis}, \citenamefont
  {Distel}, \citenamefont {Lande}, \citenamefont {Lee}, \citenamefont
  {Wildenhain},\ and\ \citenamefont {Ullman}}]{Cleveland:1998nv}%
  \BibitemOpen
  \bibfield  {author} {\bibinfo {author} {\bibfnamefont {B.~T.}\ \bibnamefont
  {Cleveland}}, \emph {et~al.},\ }\href {https://doi.org/10.1086/305343}
  {\bibfield  {journal} {\bibinfo  {journal} {Astrophys. J.}\ }\textbf
  {\bibinfo {volume} {496}},\ \bibinfo {pages} {505} (\bibinfo {year}
  {1998})}\BibitemShut {NoStop}%
\bibitem [{\citenamefont {Kaether}\ \emph {et~al.}(2010)\citenamefont
  {Kaether}, \citenamefont {Hampel}, \citenamefont {Heusser}, \citenamefont
  {Kiko},\ and\ \citenamefont {Kirsten}}]{Kaether:2010ag}%
  \BibitemOpen
  \bibfield  {author} {\bibinfo {author} {\bibfnamefont {F.}~\bibnamefont
  {Kaether}}, \bibinfo {author} {\bibfnamefont {W.}~\bibnamefont {Hampel}},
  \bibinfo {author} {\bibfnamefont {G.}~\bibnamefont {Heusser}}, \bibinfo
  {author} {\bibfnamefont {J.}~\bibnamefont {Kiko}},\ and\ \bibinfo {author}
  {\bibfnamefont {T.}~\bibnamefont {Kirsten}},\ }\href
  {https://doi.org/10.1016/j.physletb.2010.01.030} {\bibfield  {journal}
  {\bibinfo  {journal} {\plb}\ }\textbf {\bibinfo {volume} {685}},\ \bibinfo
  {pages} {47} (\bibinfo {year} {2010})},\ \Eprint
  {https://arxiv.org/abs/1001.2731} {arXiv:1001.2731} \BibitemShut {NoStop}%
\bibitem [{\citenamefont {Abdurashitov}\ \emph {et~al.}(2009)\citenamefont
  {Abdurashitov} \emph {et~al.}}]{Abdurashitov:2009tn}%
  \BibitemOpen
  \bibfield  {author} {\bibinfo {author} {\bibfnamefont {J.~N.}\ \bibnamefont
  {Abdurashitov}} \emph {et~al.} (\bibinfo {collaboration} {SAGE}),\ }\href
  {https://doi.org/10.1103/PhysRevC.80.015807} {\bibfield  {journal} {\bibinfo
  {journal} {Phys. Rev.}\ }\textbf {\bibinfo {volume} {C80}},\ \bibinfo {pages}
  {015807} (\bibinfo {year} {2009})},\ \Eprint
  {https://arxiv.org/abs/0901.2200} {arXiv:0901.2200} \BibitemShut {NoStop}%
\bibitem [{\citenamefont {Aharmim}\ \emph {et~al.}(2013)\citenamefont {Aharmim}
  \emph {et~al.}}]{Aharmim:2011vm}%
  \BibitemOpen
  \bibfield  {author} {\bibinfo {author} {\bibfnamefont {B.}~\bibnamefont
  {Aharmim}} \emph {et~al.} (\bibinfo {collaboration} {SNO}),\ }\href
  {https://doi.org/10.1103/PhysRevC.88.025501} {\bibfield  {journal} {\bibinfo
  {journal} {Phys. Rev.}\ }\textbf {\bibinfo {volume} {C88}},\ \bibinfo {pages}
  {025501} (\bibinfo {year} {2013})},\ \Eprint
  {https://arxiv.org/abs/1109.0763} {arXiv:1109.0763} \BibitemShut {NoStop}%
\bibitem [{\citenamefont {Hosaka}\ \emph {et~al.}(2006)\citenamefont {Hosaka}
  \emph {et~al.}}]{Hosaka:2005um}%
  \BibitemOpen
  \bibfield  {author} {\bibinfo {author} {\bibfnamefont {J.}~\bibnamefont
  {Hosaka}} \emph {et~al.} (\bibinfo {collaboration} {Super-Kamiokande}),\
  }\href {https://doi.org/10.1103/PhysRevD.73.112001} {\bibfield  {journal}
  {\bibinfo  {journal} {\prd}\ }\textbf {\bibinfo {volume} {73}},\ \bibinfo
  {pages} {112001} (\bibinfo {year} {2006})},\ \Eprint
  {https://arxiv.org/abs/hep-ex/0508053} {arXiv:hep-ex/0508053} \BibitemShut
  {NoStop}%
\bibitem [{\citenamefont {Cravens}\ \emph {et~al.}(2008)\citenamefont {Cravens}
  \emph {et~al.}}]{Cravens:2008aa}%
  \BibitemOpen
  \bibfield  {author} {\bibinfo {author} {\bibfnamefont {J.~P.}\ \bibnamefont
  {Cravens}} \emph {et~al.} (\bibinfo {collaboration} {Super-Kamiokande}),\
  }\href {https://doi.org/10.1103/PhysRevD.78.032002} {\bibfield  {journal}
  {\bibinfo  {journal} {\prd}\ }\textbf {\bibinfo {volume} {78}},\ \bibinfo
  {pages} {032002} (\bibinfo {year} {2008})},\ \Eprint
  {https://arxiv.org/abs/0803.4312} {arXiv:0803.4312} \BibitemShut {NoStop}%
\bibitem [{\citenamefont {Abe}\ \emph {et~al.}(2011)\citenamefont {Abe} \emph
  {et~al.}}]{Abe:2010hy}%
  \BibitemOpen
  \bibfield  {author} {\bibinfo {author} {\bibfnamefont {K.}~\bibnamefont
  {Abe}} \emph {et~al.} (\bibinfo {collaboration} {Super-Kamiokande}),\ }\href
  {https://doi.org/10.1103/PhysRevD.83.052010} {\bibfield  {journal} {\bibinfo
  {journal} {\prd}\ }\textbf {\bibinfo {volume} {83}},\ \bibinfo {pages}
  {052010} (\bibinfo {year} {2011})},\ \Eprint
  {https://arxiv.org/abs/1010.0118} {arXiv:1010.0118} \BibitemShut {NoStop}%
\bibitem [{\citenamefont {Ikeda}(2018)}]{skiv}%
  \BibitemOpen
  \bibfield  {author} {\bibinfo {author} {\bibfnamefont {M.}~\bibnamefont
  {Ikeda}},\ }\href {https://doi.org/10.5281/zenodo.1286858} {\bibinfo {title}
  {Solar neutrino measurements with super-kamiokande}} (\bibinfo {year}
  {2018}),\ \bibinfo {note} {\textit{Neutrino 2018}, Heidelberg, June 5,
  2018.}\BibitemShut {Stop}%
\bibitem [{\citenamefont {Bellini}\ \emph {et~al.}(2011)\citenamefont {Bellini}
  \emph {et~al.}}]{Bellini:2011rx}%
  \BibitemOpen
  \bibfield  {author} {\bibinfo {author} {\bibfnamefont {G.}~\bibnamefont
  {Bellini}} \emph {et~al.},\ }\href
  {https://doi.org/10.1103/PhysRevLett.107.141302} {\bibfield  {journal}
  {\bibinfo  {journal} {\prl}\ }\textbf {\bibinfo {volume} {107}},\ \bibinfo
  {pages} {141302} (\bibinfo {year} {2011})},\ \Eprint
  {https://arxiv.org/abs/1104.1816} {arXiv:1104.1816} \BibitemShut {NoStop}%
\bibitem [{\citenamefont {Bellini}\ \emph {et~al.}(2010)\citenamefont {Bellini}
  \emph {et~al.}}]{Bellini:2008mr}%
  \BibitemOpen
  \bibfield  {author} {\bibinfo {author} {\bibfnamefont {G.}~\bibnamefont
  {Bellini}} \emph {et~al.} (\bibinfo {collaboration} {Borexino}),\ }\href
  {https://doi.org/10.1103/PhysRevD.82.033006} {\bibfield  {journal} {\bibinfo
  {journal} {\prd}\ }\textbf {\bibinfo {volume} {82}},\ \bibinfo {pages}
  {033006} (\bibinfo {year} {2010})},\ \Eprint
  {https://arxiv.org/abs/0808.2868} {arXiv:0808.2868} \BibitemShut {NoStop}%
\bibitem [{\citenamefont {Bellini}\ \emph {et~al.}(2014)\citenamefont {Bellini}
  \emph {et~al.}}]{Bellini:2014uqa}%
  \BibitemOpen
  \bibfield  {author} {\bibinfo {author} {\bibfnamefont {G.}~\bibnamefont
  {Bellini}} \emph {et~al.} (\bibinfo {collaboration} {Borexino}),\ }\href
  {https://doi.org/10.1038/nature13702} {\bibfield  {journal} {\bibinfo
  {journal} {Nature}\ }\textbf {\bibinfo {volume} {512}},\ \bibinfo {pages}
  {383} (\bibinfo {year} {2014})}\BibitemShut {NoStop}%
\bibitem [{\citenamefont {Aartsen}\ \emph {et~al.}(2015)\citenamefont {Aartsen}
  \emph {et~al.}}]{Aartsen:2014yll}%
  \BibitemOpen
  \bibfield  {author} {\bibinfo {author} {\bibfnamefont {M.~G.}\ \bibnamefont
  {Aartsen}} \emph {et~al.} (\bibinfo {collaboration} {IceCube}),\ }\href
  {https://doi.org/10.1103/PhysRevD.91.072004} {\bibfield  {journal} {\bibinfo
  {journal} {\prd}\ }\textbf {\bibinfo {volume} {91}},\ \bibinfo {pages}
  {072004} (\bibinfo {year} {2015})},\ \Eprint
  {https://arxiv.org/abs/1410.7227} {arXiv:1410.7227} \BibitemShut {NoStop}%
\bibitem [{\citenamefont {Abe}\ \emph {et~al.}(2018)\citenamefont {Abe} \emph
  {et~al.}}]{Abe:2017aap}%
  \BibitemOpen
  \bibfield  {author} {\bibinfo {author} {\bibfnamefont {K.}~\bibnamefont
  {Abe}} \emph {et~al.} (\bibinfo {collaboration} {Super-Kamiokande}),\ }\href
  {https://doi.org/10.1103/PhysRevD.97.072001} {\bibfield  {journal} {\bibinfo
  {journal} {\prd}\ }\textbf {\bibinfo {volume} {97}},\ \bibinfo {pages}
  {072001} (\bibinfo {year} {2018})},\ \Eprint
  {https://arxiv.org/abs/1710.09126} {arXiv:1710.09126} \BibitemShut {NoStop}%
\bibitem [{\citenamefont {Wong}(2011)}]{Wong:2011ip}%
  \BibitemOpen
  \bibfield  {author} {\bibinfo {author} {\bibfnamefont {Y.~Y.}\ \bibnamefont
  {Wong}},\ }\href {https://doi.org/10.1146/annurev-nucl-102010-130252}
  {\bibfield  {journal} {\bibinfo  {journal} {\arnps}\ }\textbf {\bibinfo
  {volume} {61}},\ \bibinfo {pages} {69} (\bibinfo {year} {2011})},\ \Eprint
  {https://arxiv.org/abs/1111.1436} {arXiv:1111.1436} \BibitemShut {NoStop}%
\bibitem [{\citenamefont {Lesgourgues}\ and\ \citenamefont
  {Pastor}(2014)}]{Lesgourgues:2014zoa}%
  \BibitemOpen
  \bibfield  {author} {\bibinfo {author} {\bibfnamefont {J.}~\bibnamefont
  {Lesgourgues}}\ and\ \bibinfo {author} {\bibfnamefont {S.}~\bibnamefont
  {Pastor}},\ }\href {https://doi.org/10.1088/1367-2630/16/6/065002} {\bibfield
   {journal} {\bibinfo  {journal} {\njp}\ }\textbf {\bibinfo {volume} {16}},\
  \bibinfo {pages} {065002} (\bibinfo {year} {2014})},\ \Eprint
  {https://arxiv.org/abs/1404.1740} {arXiv:1404.1740} \BibitemShut {NoStop}%
\bibitem [{\citenamefont {Vagnozzi}\ \emph {et~al.}(2017)\citenamefont
  {Vagnozzi}, \citenamefont {Giusarma}, \citenamefont {Mena}, \citenamefont
  {Freese}, \citenamefont {Gerbino}, \citenamefont {Ho},\ and\ \citenamefont
  {Lattanzi}}]{Vagnozzi:2017ovm}%
  \BibitemOpen
  \bibfield  {author} {\bibinfo {author} {\bibfnamefont {S.}~\bibnamefont
  {Vagnozzi}}, \emph {et~al.},\ }\href
  {https://doi.org/10.1103/PhysRevD.96.123503} {\bibfield  {journal} {\bibinfo
  {journal} {\prd}\ }\textbf {\bibinfo {volume} {96}},\ \bibinfo {pages}
  {123503} (\bibinfo {year} {2017})},\ \Eprint
  {https://arxiv.org/abs/1701.08172} {arXiv:1701.08172} \BibitemShut {NoStop}%
\bibitem [{\citenamefont {Aghanim}\ \emph
  {et~al.}(2020{\natexlab{a}})\citenamefont {Aghanim} \emph
  {et~al.}}]{Aghanim:2018eyx}%
  \BibitemOpen
  \bibfield  {author} {\bibinfo {author} {\bibfnamefont {N.}~\bibnamefont
  {Aghanim}} \emph {et~al.} (\bibinfo {collaboration} {Planck}),\ }\href
  {https://doi.org/10.1051/0004-6361/201833910} {\bibfield  {journal} {\bibinfo
   {journal} {Astron. Astrophys.}\ }\textbf {\bibinfo {volume} {641}},\
  \bibinfo {pages} {A6} (\bibinfo {year} {2020}{\natexlab{a}})},\ \Eprint
  {https://arxiv.org/abs/1807.06209} {arXiv:1807.06209} \BibitemShut {NoStop}%
\bibitem [{\citenamefont {Loureiro}\ \emph {et~al.}(2019)\citenamefont
  {Loureiro} \emph {et~al.}}]{Loureiro:2018pdz}%
  \BibitemOpen
  \bibfield  {author} {\bibinfo {author} {\bibfnamefont {A.}~\bibnamefont
  {Loureiro}} \emph {et~al.},\ }\href
  {https://doi.org/10.1103/PhysRevLett.123.081301} {\bibfield  {journal}
  {\bibinfo  {journal} {\prl}\ }\textbf {\bibinfo {volume} {123}},\ \bibinfo
  {pages} {081301} (\bibinfo {year} {2019})},\ \Eprint
  {https://arxiv.org/abs/1811.02578} {arXiv:1811.02578} \BibitemShut {NoStop}%
\bibitem [{\citenamefont {Ivanov}\ \emph {et~al.}(2020)\citenamefont {Ivanov},
  \citenamefont {Simonovi\'c},\ and\ \citenamefont
  {Zaldarriaga}}]{Ivanov:2019hqk}%
  \BibitemOpen
  \bibfield  {author} {\bibinfo {author} {\bibfnamefont {M.~M.}\ \bibnamefont
  {Ivanov}}, \bibinfo {author} {\bibfnamefont {M.}~\bibnamefont
  {Simonovi\'c}},\ and\ \bibinfo {author} {\bibfnamefont {M.}~\bibnamefont
  {Zaldarriaga}},\ }\href {https://doi.org/10.1103/PhysRevD.101.083504}
  {\bibfield  {journal} {\bibinfo  {journal} {\prd}\ }\textbf {\bibinfo
  {volume} {101}},\ \bibinfo {pages} {083504} (\bibinfo {year} {2020})},\
  \Eprint {https://arxiv.org/abs/1912.08208} {arXiv:1912.08208} \BibitemShut
  {NoStop}%
\bibitem [{\citenamefont {Archidiacono}\ \emph {et~al.}(2020)\citenamefont
  {Archidiacono}, \citenamefont {Hannestad},\ and\ \citenamefont
  {Lesgourgues}}]{Archidiacono:2020dvx}%
  \BibitemOpen
  \bibfield  {author} {\bibinfo {author} {\bibfnamefont {M.}~\bibnamefont
  {Archidiacono}}, \bibinfo {author} {\bibfnamefont {S.}~\bibnamefont
  {Hannestad}},\ and\ \bibinfo {author} {\bibfnamefont {J.}~\bibnamefont
  {Lesgourgues}},\ }\href {https://doi.org/10.1088/1475-7516/2020/09/021}
  {\bibfield  {journal} {\bibinfo  {journal} {\jcap}\ }\textbf {\bibinfo
  {volume} {09}},\ \bibinfo {pages} {021} (\bibinfo {year} {2020})},\ \Eprint
  {https://arxiv.org/abs/2003.03354} {arXiv:2003.03354} \BibitemShut {NoStop}%
\bibitem [{\citenamefont {{Renk}}\ \emph {et~al.}(2021)\citenamefont {{Renk}},
  \citenamefont {{St\"ocker}}, \citenamefont {{Bloor}}, \citenamefont
  {{Hotinli}}, \citenamefont {{Bal{\'a}zs}}, \citenamefont {{Bringmann}},
  \citenamefont {{Gonzalo}}, \citenamefont {{Handley}}, \citenamefont {{Hoof}},
  \citenamefont {{Howlett}}, \citenamefont {{Kahlhoefer}}, \citenamefont
  {{Scott}}, \citenamefont {{Vincent}},\ and\ \citenamefont
  {{White}}}]{cosmobit}%
  \BibitemOpen
  \bibfield  {author} {\bibinfo {author} {\bibfnamefont {J.~J.}\ \bibnamefont
  {{Renk}}}, \emph {et~al.} (\bibinfo {collaboration} {\GB Cosmology
  Workgroup}),\ }\href {https://doi.org/10.1088/1475-7516/2021/02/022}
  {\bibfield  {journal} {\bibinfo  {journal} {\jcap}\ }\textbf {\bibinfo
  {volume} {02}},\ \bibinfo {pages} {022} (\bibinfo {year} {2021})},\ \Eprint
  {https://arxiv.org/abs/2009.03286} {arXiv:2009.03286} \BibitemShut {NoStop}%
\bibitem [{\citenamefont {{Athron}}\ \emph {et~al.}(2017)\citenamefont
  {{Athron}}, \citenamefont {{Bal{\'a}zs}}, \citenamefont {{Bringmann}},
  \citenamefont {{Buckley}}, \citenamefont {{Chrz{\c a}szcz}}, \citenamefont
  {{Conrad}}, \citenamefont {{Cornell}}, \citenamefont {{Dal}}, \citenamefont
  {{Dickinson}}, \citenamefont {{Edsj{\"o}}}, \citenamefont {{Farmer}},
  \citenamefont {{Gonzalo}}, \citenamefont {{Jackson}}, \citenamefont
  {{Krislock}}, \citenamefont {{Kvellestad}}, \citenamefont {{Lundberg}},
  \citenamefont {{McKay}}, \citenamefont {{Mahmoudi}}, \citenamefont
  {{Martinez}}, \citenamefont {{Putze}}, \citenamefont {{Raklev}},
  \citenamefont {{Ripken}}, \citenamefont {{Rogan}}, \citenamefont
  {{Saavedra}}, \citenamefont {{Savage}}, \citenamefont {{Scott}},
  \citenamefont {{Seo}}, \citenamefont {{Serra}}, \citenamefont {{Weniger}},
  \citenamefont {{White}},\ and\ \citenamefont {{Wild}}}]{gambit}%
  \BibitemOpen
  \bibfield  {author} {\bibinfo {author} {\bibfnamefont {P.}~\bibnamefont
  {{Athron}}}, \emph {et~al.} (\bibinfo {collaboration} {\GB Collaboration}),\
  }\href@noop {} {\bibfield  {journal} {\bibinfo  {journal} {\epjc}\ }\textbf
  {\bibinfo {volume} {77}},\ \bibinfo {pages} {784} (\bibinfo {year} {2017})},\
  \bibinfo {note} {addendum in \cite{gambit_addendum}},\ \Eprint
  {https://arxiv.org/abs/1705.07908} {arXiv:1705.07908} \BibitemShut {NoStop}%
\bibitem [{\citenamefont {Aghanim}\ \emph
  {et~al.}(2020{\natexlab{b}})\citenamefont {Aghanim} \emph
  {et~al.}}]{Aghanim:2019ame}%
  \BibitemOpen
  \bibfield  {author} {\bibinfo {author} {\bibfnamefont {N.}~\bibnamefont
  {Aghanim}} \emph {et~al.} (\bibinfo {collaboration} {Planck}),\ }\href
  {https://doi.org/10.1051/0004-6361/201936386} {\bibfield  {journal} {\bibinfo
   {journal} {Astron. Astrophys.}\ }\textbf {\bibinfo {volume} {641}},\
  \bibinfo {pages} {A5} (\bibinfo {year} {2020}{\natexlab{b}})},\ \Eprint
  {https://arxiv.org/abs/1907.12875} {arXiv:1907.12875} \BibitemShut {NoStop}%
\bibitem [{\citenamefont {Esteban}\ \emph {et~al.}(2019)\citenamefont
  {Esteban}, \citenamefont {Gonzalez-Garcia}, \citenamefont
  {Hernandez-Cabezudo}, \citenamefont {Maltoni},\ and\ \citenamefont
  {Schwetz}}]{Esteban:2018azc}%
  \BibitemOpen
  \bibfield  {author} {\bibinfo {author} {\bibfnamefont {I.}~\bibnamefont
  {Esteban}}, \bibinfo {author} {\bibfnamefont {M.}~\bibnamefont
  {Gonzalez-Garcia}}, \bibinfo {author} {\bibfnamefont {A.}~\bibnamefont
  {Hernandez-Cabezudo}}, \bibinfo {author} {\bibfnamefont {M.}~\bibnamefont
  {Maltoni}},\ and\ \bibinfo {author} {\bibfnamefont {T.}~\bibnamefont
  {Schwetz}},\ }\href {https://doi.org/10.1007/JHEP01(2019)106} {\bibfield
  {journal} {\bibinfo  {journal} {\jhep}\ }\textbf {\bibinfo {volume} {01}},\
  \bibinfo {pages} {106} (\bibinfo {year} {2019})},\ \bibinfo {note}
  {\textsf{v4.1} \href{www.nu-fit.org}{www.nu-fit.org}},\ \Eprint
  {https://arxiv.org/abs/1811.05487} {arXiv:1811.05487} \BibitemShut {NoStop}%
\bibitem [{\citenamefont {{Beutler}}\ \emph {et~al.}(2011)\citenamefont
  {{Beutler}}, \citenamefont {{Blake}}, \citenamefont {{Colless}},
  \citenamefont {{Jones}}, \citenamefont {{Staveley-Smith}}, \citenamefont
  {{Campbell}}, \citenamefont {{Parker}}, \citenamefont {{Saunders}},\ and\
  \citenamefont {{Watson}}}]{2011MNRAS.416.3017B}%
  \BibitemOpen
  \bibfield  {author} {\bibinfo {author} {\bibfnamefont {F.}~\bibnamefont
  {{Beutler}}}, \emph {et~al.},\ }\href
  {https://doi.org/10.1111/j.1365-2966.2011.19250.x} {\bibfield  {journal}
  {\bibinfo  {journal} {\mnras}\ }\textbf {\bibinfo {volume} {416}},\ \bibinfo
  {pages} {3017} (\bibinfo {year} {2011})},\ \Eprint
  {https://arxiv.org/abs/1106.3366} {arXiv:1106.3366} \BibitemShut {NoStop}%
\bibitem [{\citenamefont {{Ross}}\ \emph {et~al.}(2015)\citenamefont {{Ross}},
  \citenamefont {{Samushia}}, \citenamefont {{Howlett}}, \citenamefont
  {{Percival}}, \citenamefont {{Burden}},\ and\ \citenamefont
  {{Manera}}}]{2015MNRAS.449..835R}%
  \BibitemOpen
  \bibfield  {author} {\bibinfo {author} {\bibfnamefont {A.~J.}\ \bibnamefont
  {{Ross}}}, \emph {et~al.},\ }\href {https://doi.org/10.1093/mnras/stv154}
  {\bibfield  {journal} {\bibinfo  {journal} {\mnras}\ }\textbf {\bibinfo
  {volume} {449}},\ \bibinfo {pages} {835} (\bibinfo {year} {2015})},\ \Eprint
  {https://arxiv.org/abs/1409.3242} {arXiv:1409.3242} \BibitemShut {NoStop}%
\bibitem [{\citenamefont {Alam}\ \emph {et~al.}(2017)\citenamefont {Alam} \emph
  {et~al.}}]{Alam:2016hwk}%
  \BibitemOpen
  \bibfield  {author} {\bibinfo {author} {\bibfnamefont {S.}~\bibnamefont
  {Alam}} \emph {et~al.} (\bibinfo {collaboration} {BOSS}),\ }\href
  {https://doi.org/10.1093/mnras/stx721} {\bibfield  {journal} {\bibinfo
  {journal} {\mnras}\ }\textbf {\bibinfo {volume} {470}},\ \bibinfo {pages}
  {2617} (\bibinfo {year} {2017})},\ \Eprint {https://arxiv.org/abs/1607.03155}
  {arXiv:1607.03155} \BibitemShut {NoStop}%
\bibitem [{\citenamefont {Ata}\ \emph {et~al.}(2018)\citenamefont {Ata} \emph
  {et~al.}}]{Ata:2017dya}%
  \BibitemOpen
  \bibfield  {author} {\bibinfo {author} {\bibfnamefont {M.}~\bibnamefont
  {Ata}} \emph {et~al.},\ }\href {https://doi.org/10.1093/mnras/stx2630}
  {\bibfield  {journal} {\bibinfo  {journal} {Mon. Not. Roy. Astron. Soc.}\
  }\textbf {\bibinfo {volume} {473}},\ \bibinfo {pages} {4773} (\bibinfo {year}
  {2018})},\ \Eprint {https://arxiv.org/abs/1705.06373} {arXiv:1705.06373}
  \BibitemShut {NoStop}%
\bibitem [{\citenamefont {{Bautista}}\ \emph {et~al.}(2018)\citenamefont
  {{Bautista}}, \citenamefont {{Vargas-Maga{\~n}a}}, \citenamefont {{Dawson}},
  \citenamefont {{Percival}}, \citenamefont {{Brinkmann}}, \citenamefont
  {{Brownstein}}, \citenamefont {{Camacho}}, \citenamefont {{Comparat}},
  \citenamefont {{Gil-Mar{\'\i}n}}, \citenamefont {{Mueller}}, \citenamefont
  {{Newman}}, \citenamefont {{Prakash}}, \citenamefont {{Ross}}, \citenamefont
  {{Schneider}}, \citenamefont {{Seo}}, \citenamefont {{Tinker}}, \citenamefont
  {{Tojeiro}}, \citenamefont {{Zhai}},\ and\ \citenamefont
  {{Zhao}}}]{Bautista_2018}%
  \BibitemOpen
  \bibfield  {author} {\bibinfo {author} {\bibfnamefont {J.~E.}\ \bibnamefont
  {{Bautista}}}, \emph {et~al.},\ }\href
  {https://doi.org/10.3847/1538-4357/aacea5} {\bibfield  {journal} {\bibinfo
  {journal} {\apj}\ }\textbf {\bibinfo {volume} {863}},\ \bibinfo {eid} {110}
  (\bibinfo {year} {2018})},\ \Eprint {https://arxiv.org/abs/1712.08064}
  {arXiv:1712.08064} \BibitemShut {NoStop}%
\bibitem [{\citenamefont {{Blomqvist}}\ \emph {et~al.}(2019)\citenamefont
  {{Blomqvist}}, \citenamefont {{du Mas des Bourboux}}, \citenamefont
  {{Busca}}, \citenamefont {{de Sainte Agathe}}, \citenamefont {{Rich}},
  \citenamefont {{Balland}}, \citenamefont {{Bautista}}, \citenamefont
  {{Dawson}}, \citenamefont {{Font-Ribera}}, \citenamefont {{Guy}},
  \citenamefont {{Le Goff}}, \citenamefont {{Palanque-Delabrouille}},
  \citenamefont {{Percival}}, \citenamefont {{P{\'e}rez-R{\`a}fols}},
  \citenamefont {{Pieri}}, \citenamefont {{Schneider}}, \citenamefont
  {{Slosar}},\ and\ \citenamefont {{Y{\`e}che}}}]{Blomqvist2019}%
  \BibitemOpen
  \bibfield  {author} {\bibinfo {author} {\bibfnamefont {M.}~\bibnamefont
  {{Blomqvist}}}, \emph {et~al.},\ }\href
  {https://doi.org/10.1051/0004-6361/201935641} {\bibfield  {journal} {\bibinfo
   {journal} {\aap}\ }\textbf {\bibinfo {volume} {629}},\ \bibinfo {eid} {A86}
  (\bibinfo {year} {2019})},\ \Eprint {https://arxiv.org/abs/1904.03430}
  {arXiv:1904.03430} \BibitemShut {NoStop}%
\bibitem [{\citenamefont {{Abbott}}\ \emph {et~al.}(2019)\citenamefont
  {{Abbott}}, \citenamefont {{Abdalla}}, \citenamefont {{Alarcon}},
  \citenamefont {{Allam}}, \citenamefont {{Andrade-Oliveira}}, \citenamefont
  {{Annis}}, \citenamefont {{Avila}}, \citenamefont {{Banerji}}, \citenamefont
  {{Banik}}, \citenamefont {{Bechtol}}, \citenamefont {{Bernstein}},
  \citenamefont {{Bernstein}}, \citenamefont {{Bertin}}, \citenamefont
  {{Brooks}}, \citenamefont {{Buckley-Geer}}, \citenamefont {{Burke}},
  \citenamefont {{Camacho}}, \citenamefont {{Carnero Rosell}}, \citenamefont
  {{Carrasco Kind}}, \citenamefont {{Carretero}}, \citenamefont {{Castander}},
  \citenamefont {{Cawthon}}, \citenamefont {{Chan}}, \citenamefont {{Crocce}},
  \citenamefont {{Cunha}}, \citenamefont {{D'Andrea}}, \citenamefont {{da
  Costa}}, \citenamefont {{Davis}}, \citenamefont {{De Vicente}}, \citenamefont
  {{DePoy}}, \citenamefont {{Desai}}, \citenamefont {{Diehl}}, \citenamefont
  {{Doel}}, \citenamefont {{Drlica-Wagner}}, \citenamefont {{Eifler}},
  \citenamefont {{Elvin-Poole}}, \citenamefont {{Estrada}}, \citenamefont
  {{Evrard}}, \citenamefont {{Flaugher}}, \citenamefont {{Fosalba}},
  \citenamefont {{Frieman}}, \citenamefont {{Garc{\'\i}a-Bellido}},
  \citenamefont {{Gaztanaga}}, \citenamefont {{Gerdes}}, \citenamefont
  {{Giannantonio}}, \citenamefont {{Gruen}}, \citenamefont {{Gruendl}},
  \citenamefont {{Gschwend}}, \citenamefont {{Gutierrez}}, \citenamefont
  {{Hartley}}, \citenamefont {{Hollowood}}, \citenamefont {{Honscheid}},
  \citenamefont {{Hoyle}}, \citenamefont {{Jain}}, \citenamefont {{James}},
  \citenamefont {{Jeltema}}, \citenamefont {{Johnson}}, \citenamefont {{Kent}},
  \citenamefont {{Kokron}}, \citenamefont {{Krause}}, \citenamefont {{Kuehn}},
  \citenamefont {{Kuhlmann}}, \citenamefont {{Kuropatkin}}, \citenamefont
  {{Lacasa}}, \citenamefont {{Lahav}}, \citenamefont {{Lima}}, \citenamefont
  {{Lin}}, \citenamefont {{Maia}}, \citenamefont {{Manera}}, \citenamefont
  {{Marriner}}, \citenamefont {{Marshall}}, \citenamefont {{Martini}},
  \citenamefont {{Melchior}}, \citenamefont {{Menanteau}}, \citenamefont
  {{Miller}}, \citenamefont {{Miquel}}, \citenamefont {{Mohr}}, \citenamefont
  {{Neilsen}}, \citenamefont {{Percival}}, \citenamefont {{Plazas}},
  \citenamefont {{Porredon}}, \citenamefont {{Romer}}, \citenamefont
  {{Roodman}}, \citenamefont {{Rosenfeld}}, \citenamefont {{Ross}},
  \citenamefont {{Rozo}}, \citenamefont {{Rykoff}}, \citenamefont {{Sako}},
  \citenamefont {{Sanchez}}, \citenamefont {{Santiago}}, \citenamefont
  {{Scarpine}}, \citenamefont {{Schindler}}, \citenamefont {{Schubnell}},
  \citenamefont {{Serrano}}, \citenamefont {{Sevilla-Noarbe}}, \citenamefont
  {{Sheldon}}, \citenamefont {{Smith}}, \citenamefont {{Smith}}, \citenamefont
  {{Sobreira}}, \citenamefont {{Suchyta}}, \citenamefont {{Swanson}},
  \citenamefont {{Tarle}}, \citenamefont {{Thomas}}, \citenamefont {{Troxel}},
  \citenamefont {{Tucker}}, \citenamefont {{Vikram}}, \citenamefont {{Walker}},
  \citenamefont {{Wechsler}}, \citenamefont {{Weller}}, \citenamefont
  {{Yanny}},\ and\ \citenamefont {{Zhang}}}]{Abbott_2018}%
  \BibitemOpen
  \bibfield  {author} {\bibinfo {author} {\bibfnamefont {T.~M.~C.}\
  \bibnamefont {{Abbott}}}, \emph {et~al.},\ }\href
  {https://doi.org/10.1093/mnras/sty3351} {\bibfield  {journal} {\bibinfo
  {journal} {\mnras}\ }\textbf {\bibinfo {volume} {483}},\ \bibinfo {pages}
  {4866} (\bibinfo {year} {2019})},\ \Eprint {https://arxiv.org/abs/1712.06209}
  {arXiv:1712.06209} \BibitemShut {NoStop}%
\bibitem [{\citenamefont {Riess}\ \emph {et~al.}(2019)\citenamefont {Riess},
  \citenamefont {Casertano}, \citenamefont {Yuan}, \citenamefont {Macri},\ and\
  \citenamefont {Scolnic}}]{Riess:2019cxk}%
  \BibitemOpen
  \bibfield  {author} {\bibinfo {author} {\bibfnamefont {A.~G.}\ \bibnamefont
  {Riess}}, \bibinfo {author} {\bibfnamefont {S.}~\bibnamefont {Casertano}},
  \bibinfo {author} {\bibfnamefont {W.}~\bibnamefont {Yuan}}, \bibinfo {author}
  {\bibfnamefont {L.~M.}\ \bibnamefont {Macri}},\ and\ \bibinfo {author}
  {\bibfnamefont {D.}~\bibnamefont {Scolnic}},\ }\href
  {https://doi.org/10.3847/1538-4357/ab1422} {\bibfield  {journal} {\bibinfo
  {journal} {Astrophys. J.}\ }\textbf {\bibinfo {volume} {876}},\ \bibinfo
  {pages} {85} (\bibinfo {year} {2019})},\ \Eprint
  {https://arxiv.org/abs/1903.07603} {arXiv:1903.07603} \BibitemShut {NoStop}%
\bibitem [{\citenamefont {{Wong}}\ \emph {et~al.}(2020)\citenamefont {{Wong}},
  \citenamefont {{Suyu}}, \citenamefont {{Chen}}, \citenamefont {{Rusu}},
  \citenamefont {{Millon}}, \citenamefont {{Sluse}}, \citenamefont {{Bonvin}},
  \citenamefont {{Fassnacht}}, \citenamefont {{Taubenberger}}, \citenamefont
  {{Auger}}, \citenamefont {{Birrer}}, \citenamefont {{Chan}}, \citenamefont
  {{Courbin}}, \citenamefont {{Hilbert}}, \citenamefont {{Tihhonova}},
  \citenamefont {{Treu}}, \citenamefont {{Agnello}}, \citenamefont {{Ding}},
  \citenamefont {{Jee}}, \citenamefont {{Komatsu}}, \citenamefont {{Shajib}},
  \citenamefont {{Sonnenfeld}}, \citenamefont {{Bland ford}}, \citenamefont
  {{Koopmans}}, \citenamefont {{Marshall}},\ and\ \citenamefont
  {{Meylan}}}]{2019arXiv190704869W}%
  \BibitemOpen
  \bibfield  {author} {\bibinfo {author} {\bibfnamefont {K.~C.}\ \bibnamefont
  {{Wong}}}, \emph {et~al.},\ }\href {https://doi.org/10.1093/mnras/stz3094}
  {\bibfield  {journal} {\bibinfo  {journal} {\mnras}\ }\textbf {\bibinfo
  {volume} {498}},\ \bibinfo {pages} {1420} (\bibinfo {year} {2020})},\ \Eprint
  {https://arxiv.org/abs/1907.04869} {arXiv:1907.04869} \BibitemShut {NoStop}%
\bibitem [{\citenamefont {{Freedman}}\ \emph {et~al.}(2019)\citenamefont
  {{Freedman}}, \citenamefont {{Madore}}, \citenamefont {{Hatt}}, \citenamefont
  {{Hoyt}}, \citenamefont {{Jang}}, \citenamefont {{Beaton}}, \citenamefont
  {{Burns}}, \citenamefont {{Lee}}, \citenamefont {{Monson}}, \citenamefont
  {{Neeley}}, \citenamefont {{Phillips}}, \citenamefont {{Rich}},\ and\
  \citenamefont {{Seibert}}}]{2019ApJ...882...34F}%
  \BibitemOpen
  \bibfield  {author} {\bibinfo {author} {\bibfnamefont {W.~L.}\ \bibnamefont
  {{Freedman}}}, \emph {et~al.},\ }\href
  {https://doi.org/10.3847/1538-4357/ab2f73} {\bibfield  {journal} {\bibinfo
  {journal} {\apj}\ }\textbf {\bibinfo {volume} {882}},\ \bibinfo {eid} {34}
  (\bibinfo {year} {2019})},\ \Eprint {https://arxiv.org/abs/1907.05922}
  {arXiv:1907.05922} \BibitemShut {NoStop}%
\bibitem [{\citenamefont {Chrzaszcz}\ \emph {et~al.}(2020)\citenamefont
  {Chrzaszcz}, \citenamefont {Drewes}, \citenamefont {Gonzalo}, \citenamefont
  {Harz}, \citenamefont {Krishnamurthy},\ and\ \citenamefont {Weniger}}]{RHN}%
  \BibitemOpen
  \bibfield  {author} {\bibinfo {author} {\bibfnamefont {M.}~\bibnamefont
  {Chrzaszcz}}, \emph {et~al.},\ }\href
  {https://doi.org/10.1140/epjc/s10052-020-8073-9} {\bibfield  {journal}
  {\bibinfo  {journal} {\epjc}\ }\textbf {\bibinfo {volume} {80}},\ \bibinfo
  {pages} {569} (\bibinfo {year} {2020})},\ \Eprint
  {https://arxiv.org/abs/1908.02302} {arXiv:1908.02302} \BibitemShut {NoStop}%
\bibitem [{\citenamefont {{Athron}}\ \emph
  {et~al.}(2018{\natexlab{a}})\citenamefont {{Athron}}, \citenamefont
  {{Bal{\'a}zs}}, \citenamefont {{Dal}}, \citenamefont {{Edsj{\"o}}},
  \citenamefont {{Farmer}}, \citenamefont {{Gonzalo}}, \citenamefont
  {{Kvellestad}}, \citenamefont {{McKay}}, \citenamefont {{Putze}},
  \citenamefont {{Rogan}}, \citenamefont {{Scott}}, \citenamefont {{Weniger}},\
  and\ \citenamefont {{White}}}]{SDPBit}%
  \BibitemOpen
  \bibfield  {author} {\bibinfo {author} {\bibfnamefont {P.}~\bibnamefont
  {{Athron}}}, \emph {et~al.} (\bibinfo {collaboration} {\GB Models
  Workgroup}),\ }\href {https://doi.org/10.1140/epjc/s10052-017-5390-8}
  {\bibfield  {journal} {\bibinfo  {journal} {\epjc}\ }\textbf {\bibinfo
  {volume} {78}},\ \bibinfo {eid} {22} (\bibinfo {year}
  {2018}{\natexlab{a}})},\ \Eprint {https://arxiv.org/abs/1705.07936}
  {arXiv:1705.07936} \BibitemShut {NoStop}%
\bibitem [{\citenamefont {Esteban}\ \emph {et~al.}(2020)\citenamefont
  {Esteban}, \citenamefont {Gonzalez-Garcia}, \citenamefont {Maltoni},
  \citenamefont {Schwetz},\ and\ \citenamefont {Zhou}}]{Esteban:2020cvm}%
  \BibitemOpen
  \bibfield  {author} {\bibinfo {author} {\bibfnamefont {I.}~\bibnamefont
  {Esteban}}, \bibinfo {author} {\bibfnamefont {M.}~\bibnamefont
  {Gonzalez-Garcia}}, \bibinfo {author} {\bibfnamefont {M.}~\bibnamefont
  {Maltoni}}, \bibinfo {author} {\bibfnamefont {T.}~\bibnamefont {Schwetz}},\
  and\ \bibinfo {author} {\bibfnamefont {A.}~\bibnamefont {Zhou}},\ }\href
  {https://doi.org/10.1007/JHEP09(2020)178} {\bibfield  {journal} {\bibinfo
  {journal} {\jhep}\ }\textbf {\bibinfo {volume} {09}},\ \bibinfo {pages} {178}
  (\bibinfo {year} {2020})},\ \Eprint {https://arxiv.org/abs/2007.14792}
  {arXiv:2007.14792} \BibitemShut {NoStop}%
\bibitem [{\citenamefont {Tanabashi}\ \emph {et~al.}(date)\citenamefont
  {Tanabashi} \emph {et~al.}}]{Tanabashi:2018oca}%
  \BibitemOpen
  \bibfield  {author} {\bibinfo {author} {\bibfnamefont {M.}~\bibnamefont
  {Tanabashi}} \emph {et~al.} (\bibinfo {collaboration} {Particle Data
  Group}),\ }\href {https://doi.org/10.1103/PhysRevD.98.030001} {\bibfield
  {journal} {\bibinfo  {journal} {\prd}\ }\textbf {\bibinfo {volume} {98}},\
  \bibinfo {pages} {030001} (\bibinfo {year} {2018 and 2019
  update})}\BibitemShut {NoStop}%
\bibitem [{\citenamefont {Cooke}\ \emph {et~al.}(2018)\citenamefont {Cooke},
  \citenamefont {Pettini},\ and\ \citenamefont {Steidel}}]{Cooke:2017cwo}%
  \BibitemOpen
  \bibfield  {author} {\bibinfo {author} {\bibfnamefont {R.~J.}\ \bibnamefont
  {Cooke}}, \bibinfo {author} {\bibfnamefont {M.}~\bibnamefont {Pettini}},\
  and\ \bibinfo {author} {\bibfnamefont {C.~C.}\ \bibnamefont {Steidel}},\
  }\href {https://doi.org/10.3847/1538-4357/aaab53} {\bibfield  {journal}
  {\bibinfo  {journal} {Astrophys. J.}\ }\textbf {\bibinfo {volume} {855}},\
  \bibinfo {pages} {102} (\bibinfo {year} {2018})},\ \Eprint
  {https://arxiv.org/abs/1710.11129} {arXiv:1710.11129} \BibitemShut {NoStop}%
\bibitem [{\citenamefont {Scolnic}\ \emph {et~al.}(2018)\citenamefont {Scolnic}
  \emph {et~al.}}]{Scolnic:2017caz}%
  \BibitemOpen
  \bibfield  {author} {\bibinfo {author} {\bibfnamefont {D.}~\bibnamefont
  {Scolnic}} \emph {et~al.},\ }\href {https://doi.org/10.3847/1538-4357/aab9bb}
  {\bibfield  {journal} {\bibinfo  {journal} {Astrophys. J.}\ }\textbf
  {\bibinfo {volume} {859}},\ \bibinfo {pages} {101} (\bibinfo {year}
  {2018})},\ \Eprint {https://arxiv.org/abs/1710.00845} {arXiv:1710.00845}
  \BibitemShut {NoStop}%
\bibitem [{\citenamefont {Carter}\ \emph {et~al.}(2018)\citenamefont {Carter},
  \citenamefont {Beutler}, \citenamefont {Percival}, \citenamefont {Blake},
  \citenamefont {Koda},\ and\ \citenamefont {Ross}}]{Carter_2018}%
  \BibitemOpen
  \bibfield  {author} {\bibinfo {author} {\bibfnamefont {P.}~\bibnamefont
  {Carter}}, \emph {et~al.},\ }\href {https://doi.org/10.1093/mnras/sty2405}
  {\bibfield  {journal} {\bibinfo  {journal} {Monthly Notices of the Royal
  Astronomical Society}\ }\textbf {\bibinfo {volume} {481}},\ \bibinfo {pages}
  {2371–2383} (\bibinfo {year} {2018})}\BibitemShut {NoStop}%
\bibitem [{\citenamefont {Seo}\ and\ \citenamefont
  {Eisenstein}(2007)}]{Seo_2007}%
  \BibitemOpen
  \bibfield  {author} {\bibinfo {author} {\bibfnamefont {H.}~\bibnamefont
  {Seo}}\ and\ \bibinfo {author} {\bibfnamefont {D.~J.}\ \bibnamefont
  {Eisenstein}},\ }\href {https://doi.org/10.1086/519549} {\bibfield  {journal}
  {\bibinfo  {journal} {The Astrophysical Journal}\ }\textbf {\bibinfo {volume}
  {665}},\ \bibinfo {pages} {14} (\bibinfo {year} {2007})}\BibitemShut
  {NoStop}%
\bibitem [{\citenamefont {Font-Ribera}\ \emph
  {et~al.}(2014{\natexlab{a}})\citenamefont {Font-Ribera}, \citenamefont
  {McDonald}, \citenamefont {Mostek}, \citenamefont {Reid}, \citenamefont
  {Seo},\ and\ \citenamefont {Slosar}}]{Font_Ribera_2014}%
  \BibitemOpen
  \bibfield  {author} {\bibinfo {author} {\bibfnamefont {A.}~\bibnamefont
  {Font-Ribera}}, \emph {et~al.},\ }\href
  {https://doi.org/10.1088/1475-7516/2014/05/023} {\bibfield  {journal}
  {\bibinfo  {journal} {Journal of Cosmology and Astroparticle Physics}\
  }\textbf {\bibinfo {volume} {2014}}\bibinfo  {number} { (05)},\ \bibinfo
  {pages} {023–023}}\BibitemShut {NoStop}%
\bibitem [{\citenamefont {Audren}\ \emph {et~al.}(2013)\citenamefont {Audren},
  \citenamefont {Lesgourgues}, \citenamefont {Benabed},\ and\ \citenamefont
  {Prunet}}]{Audren:2012wb}%
  \BibitemOpen
\bibfield  {number} {  }\bibfield  {author} {\bibinfo {author} {\bibfnamefont
  {B.}~\bibnamefont {Audren}}, \bibinfo {author} {\bibfnamefont
  {J.}~\bibnamefont {Lesgourgues}}, \bibinfo {author} {\bibfnamefont
  {K.}~\bibnamefont {Benabed}},\ and\ \bibinfo {author} {\bibfnamefont
  {S.}~\bibnamefont {Prunet}},\ }\href
  {https://doi.org/10.1088/1475-7516/2013/02/001} {\bibfield  {journal}
  {\bibinfo  {journal} {\jcap}\ }\textbf {\bibinfo {volume} {1302}},\ \bibinfo
  {pages} {001} (\bibinfo {year} {2013})},\ \Eprint
  {https://arxiv.org/abs/1210.7183} {arXiv:1210.7183} \BibitemShut {NoStop}%
\bibitem [{\citenamefont {Brinckmann}\ and\ \citenamefont
  {Lesgourgues}(2019)}]{brinckmann2018montepython}%
  \BibitemOpen
  \bibfield  {author} {\bibinfo {author} {\bibfnamefont {T.}~\bibnamefont
  {Brinckmann}}\ and\ \bibinfo {author} {\bibfnamefont {J.}~\bibnamefont
  {Lesgourgues}},\ }\href {https://doi.org/10.1016/j.dark.2018.100260}
  {\bibfield  {journal} {\bibinfo  {journal} {Phys. Dark Univ.}\ }\textbf
  {\bibinfo {volume} {24}},\ \bibinfo {pages} {100260} (\bibinfo {year}
  {2019})},\ \Eprint {https://arxiv.org/abs/1804.07261} {arXiv:1804.07261}
  \BibitemShut {NoStop}%
\bibitem [{\citenamefont {Arbey}(2012)}]{Arbey:2011nf}%
  \BibitemOpen
  \bibfield  {author} {\bibinfo {author} {\bibfnamefont {A.}~\bibnamefont
  {Arbey}},\ }\href {https://doi.org/10.1016/j.cpc.2012.03.018} {\bibfield
  {journal} {\bibinfo  {journal} {Comput. Phys. Commun.}\ }\textbf {\bibinfo
  {volume} {183}},\ \bibinfo {pages} {1822} (\bibinfo {year} {2012})},\ \Eprint
  {https://arxiv.org/abs/1106.1363} {arXiv:1106.1363} \BibitemShut {NoStop}%
\bibitem [{\citenamefont {Arbey}\ \emph {et~al.}(2020)\citenamefont {Arbey},
  \citenamefont {Auffinger}, \citenamefont {Hickerson},\ and\ \citenamefont
  {Jenssen}}]{Arbey:2018zfh}%
  \BibitemOpen
  \bibfield  {author} {\bibinfo {author} {\bibfnamefont {A.}~\bibnamefont
  {Arbey}}, \bibinfo {author} {\bibfnamefont {J.}~\bibnamefont {Auffinger}},
  \bibinfo {author} {\bibfnamefont {K.~P.}\ \bibnamefont {Hickerson}},\ and\
  \bibinfo {author} {\bibfnamefont {E.~S.}\ \bibnamefont {Jenssen}},\ }\href
  {https://doi.org/10.1016/j.cpc.2019.106982} {\bibfield  {journal} {\bibinfo
  {journal} {Comput. Phys. Commun.}\ }\textbf {\bibinfo {volume} {248}},\
  \bibinfo {pages} {106982} (\bibinfo {year} {2020})},\ \Eprint
  {https://arxiv.org/abs/1806.11095} {arXiv:1806.11095} \BibitemShut {NoStop}%
\bibitem [{\citenamefont {Blas}\ \emph {et~al.}(2011)\citenamefont {Blas},
  \citenamefont {Lesgourgues},\ and\ \citenamefont {Tram}}]{Blas:2011rf}%
  \BibitemOpen
  \bibfield  {author} {\bibinfo {author} {\bibfnamefont {D.}~\bibnamefont
  {Blas}}, \bibinfo {author} {\bibfnamefont {J.}~\bibnamefont {Lesgourgues}},\
  and\ \bibinfo {author} {\bibfnamefont {T.}~\bibnamefont {Tram}},\ }\href
  {https://doi.org/10.1088/1475-7516/2011/07/034} {\bibfield  {journal}
  {\bibinfo  {journal} {\jcap}\ }\textbf {\bibinfo {volume} {1107}},\ \bibinfo
  {pages} {034} (\bibinfo {year} {2011})},\ \Eprint
  {https://arxiv.org/abs/1104.2933} {arXiv:1104.2933} \BibitemShut {NoStop}%
\bibitem [{\citenamefont {de~Salas}\ and\ \citenamefont
  {Pastor}(2016)}]{deSalas:2016ztq}%
  \BibitemOpen
  \bibfield  {author} {\bibinfo {author} {\bibfnamefont {P.~F.}\ \bibnamefont
  {de~Salas}}\ and\ \bibinfo {author} {\bibfnamefont {S.}~\bibnamefont
  {Pastor}},\ }\href {https://doi.org/10.1088/1475-7516/2016/07/051} {\bibfield
   {journal} {\bibinfo  {journal} {JCAP}\ }\textbf {\bibinfo {volume}
  {1607}}\bibfield  {number} {\bibinfo  {number} { (07)},\ \bibinfo {pages}
  {051}},\ }\Eprint {https://arxiv.org/abs/1606.06986} {arXiv:1606.06986}
  \BibitemShut {NoStop}%
\bibitem [{\citenamefont {Akita}\ and\ \citenamefont
  {Yamaguchi}(2020)}]{Akita:2020szl}%
  \BibitemOpen
  \bibfield  {author} {\bibinfo {author} {\bibfnamefont {K.}~\bibnamefont
  {Akita}}\ and\ \bibinfo {author} {\bibfnamefont {M.}~\bibnamefont
  {Yamaguchi}},\ }\href {https://doi.org/10.1088/1475-7516/2020/08/012}
  {\bibfield  {journal} {\bibinfo  {journal} {JCAP}\ }\textbf {\bibinfo
  {volume} {08}},\ \bibinfo {pages} {012}},\ \Eprint
  {https://arxiv.org/abs/2005.07047} {arXiv:2005.07047} \BibitemShut {NoStop}%
\bibitem [{\citenamefont {Froustey}\ \emph {et~al.}(2020)\citenamefont
  {Froustey}, \citenamefont {Pitrou},\ and\ \citenamefont
  {Volpe}}]{Froustey:2020mcq}%
  \BibitemOpen
  \bibfield  {author} {\bibinfo {author} {\bibfnamefont {J.}~\bibnamefont
  {Froustey}}, \bibinfo {author} {\bibfnamefont {C.}~\bibnamefont {Pitrou}},\
  and\ \bibinfo {author} {\bibfnamefont {M.~C.}\ \bibnamefont {Volpe}},\ }\href
  {https://doi.org/10.1088/1475-7516/2020/12/015} {\bibfield  {journal}
  {\bibinfo  {journal} {\jcap}\ }\textbf {\bibinfo {volume} {12}},\ \bibinfo
  {pages} {015} (\bibinfo {year} {2020})},\ \Eprint
  {https://arxiv.org/abs/2008.01074} {arXiv:2008.01074} \BibitemShut {NoStop}%
\bibitem [{\citenamefont {Zyla}\ \emph {et~al.}(2020)\citenamefont {Zyla} \emph
  {et~al.}}]{PDG20}%
  \BibitemOpen
  \bibfield  {author} {\bibinfo {author} {\bibfnamefont {P.~A.}\ \bibnamefont
  {Zyla}} \emph {et~al.} (\bibinfo {collaboration} {Particle Data Group}),\
  }\href@noop {} {\bibfield  {journal} {\bibinfo  {journal} {Prog. Theor. Exp.
  Phys.}\ }\textbf {\bibinfo {volume} {083}},\ \bibinfo {pages} {C01} (\bibinfo
  {year} {2020})}\BibitemShut {NoStop}%
\bibitem [{\citenamefont {{Handley}}\ \emph {et~al.}(2015)\citenamefont
  {{Handley}}, \citenamefont {{Hobson}},\ and\ \citenamefont
  {{Lasenby}}}]{Handley:2015}%
  \BibitemOpen
  \bibfield  {author} {\bibinfo {author} {\bibfnamefont {W.~J.}\ \bibnamefont
  {{Handley}}}, \bibinfo {author} {\bibfnamefont {M.~P.}\ \bibnamefont
  {{Hobson}}},\ and\ \bibinfo {author} {\bibfnamefont {A.~N.}\ \bibnamefont
  {{Lasenby}}},\ }\href {https://doi.org/10.1093/mnras/stv1911} {\bibfield
  {journal} {\bibinfo  {journal} {\mnras}\ }\textbf {\bibinfo {volume} {453}},\
  \bibinfo {pages} {4384} (\bibinfo {year} {2015})},\ \Eprint
  {https://arxiv.org/abs/1506.00171} {arXiv:1506.00171} \BibitemShut {NoStop}%
\bibitem [{\citenamefont {{Martinez}}\ \emph {et~al.}(2017)\citenamefont
  {{Martinez}}, \citenamefont {{McKay}}, \citenamefont {{Farmer}},
  \citenamefont {{Scott}}, \citenamefont {{Roebber}}, \citenamefont {{Putze}},\
  and\ \citenamefont {{Conrad}}}]{scannerbit}%
  \BibitemOpen
  \bibfield  {author} {\bibinfo {author} {\bibfnamefont {G.~D.}\ \bibnamefont
  {{Martinez}}}, \emph {et~al.} (\bibinfo {collaboration} {\GB Scanner
  Workgroup}),\ }\href {https://doi.org/10.1140/epjc/s10052-017-5274-y}
  {\bibfield  {journal} {\bibinfo  {journal} {\epjc}\ }\textbf {\bibinfo
  {volume} {77}},\ \bibinfo {pages} {761} (\bibinfo {year} {2017})},\ \Eprint
  {https://arxiv.org/abs/1705.07959} {arXiv:1705.07959} \BibitemShut {NoStop}%
\bibitem [{\citenamefont {Alam}\ \emph {et~al.}(2021)\citenamefont {Alam} \emph
  {et~al.}}]{Alam:2020sor}%
  \BibitemOpen
  \bibfield  {author} {\bibinfo {author} {\bibfnamefont {S.}~\bibnamefont
  {Alam}} \emph {et~al.} (\bibinfo {collaboration} {eBOSS}),\ }\href
  {https://doi.org/10.1103/PhysRevD.103.083533} {\bibfield  {journal} {\bibinfo
   {journal} {\prd}\ }\textbf {\bibinfo {volume} {103}},\ \bibinfo {pages}
  {083533} (\bibinfo {year} {2021})},\ \Eprint
  {https://arxiv.org/abs/2007.08991} {arXiv:2007.08991} \BibitemShut {NoStop}%
\bibitem [{\citenamefont {de~Salas}\ \emph {et~al.}(2021)\citenamefont
  {de~Salas}, \citenamefont {Forero}, \citenamefont {Gariazzo}, \citenamefont
  {Mart\'\i{}nez-Mirav\'e}, \citenamefont {Mena}, \citenamefont {Ternes},
  \citenamefont {T\'ortola},\ and\ \citenamefont {Valle}}]{deSalas:2020pgw}%
  \BibitemOpen
  \bibfield  {author} {\bibinfo {author} {\bibfnamefont {P.}~\bibnamefont
  {de~Salas}}, \emph {et~al.},\ }\href
  {https://doi.org/10.1007/JHEP02(2021)071} {\bibfield  {journal} {\bibinfo
  {journal} {\jhep}\ }\textbf {\bibinfo {volume} {02}},\ \bibinfo {pages} {071}
  (\bibinfo {year} {2021})},\ \Eprint {https://arxiv.org/abs/2006.11237}
  {arXiv:2006.11237} \BibitemShut {NoStop}%
\bibitem [{\citenamefont {Aker}\ \emph {et~al.}(2019)\citenamefont {Aker} \emph
  {et~al.}}]{Aker:2019uuj}%
  \BibitemOpen
  \bibfield  {author} {\bibinfo {author} {\bibfnamefont {M.}~\bibnamefont
  {Aker}} \emph {et~al.} (\bibinfo {collaboration} {KATRIN}),\ }\href
  {https://doi.org/10.1103/PhysRevLett.123.221802} {\bibfield  {journal}
  {\bibinfo  {journal} {\prl}\ }\textbf {\bibinfo {volume} {123}},\ \bibinfo
  {pages} {221802} (\bibinfo {year} {2019})},\ \Eprint
  {https://arxiv.org/abs/1909.06048} {arXiv:1909.06048} \BibitemShut {NoStop}%
\bibitem [{\citenamefont {Font-Ribera}\ \emph
  {et~al.}(2014{\natexlab{b}})\citenamefont {Font-Ribera} \emph
  {et~al.}}]{Font-Ribera:2013wce}%
  \BibitemOpen
  \bibfield  {author} {\bibinfo {author} {\bibfnamefont {A.}~\bibnamefont
  {Font-Ribera}} \emph {et~al.} (\bibinfo {collaboration} {BOSS}),\ }\href
  {https://doi.org/10.1088/1475-7516/2014/05/027} {\bibfield  {journal}
  {\bibinfo  {journal} {\jcap}\ }\textbf {\bibinfo {volume} {1405}},\ \bibinfo
  {pages} {027} (\bibinfo {year} {2014}{\natexlab{b}})},\ \Eprint
  {https://arxiv.org/abs/1311.1767} {arXiv:1311.1767} \BibitemShut {NoStop}%
\bibitem [{Cos()}]{CosmoBit_numass_zenodo}%
  \BibitemOpen
  \href@noop {} {}\bibinfo {note} {\GB Cosmology Workgroup,
  \textit{Supplementary Data: Strengthening the bound on the mass of the
  lightest neutrino with terrestrial and cosmological experiments}, (2020),
  \href{https://doi.org/10.5281/zenodo.4005381}{\nolinkurl{https://doi.org/10.5281/zenodo.4005381}}}\BibitemShut
  {NoStop}%
\bibitem [{\citenamefont {Hunter}(2007)}]{Hunter:2007}%
  \BibitemOpen
  \bibfield  {author} {\bibinfo {author} {\bibfnamefont {J.~D.}\ \bibnamefont
  {Hunter}},\ }\href {https://doi.org/10.1109/MCSE.2007.55} {\bibfield
  {journal} {\bibinfo  {journal} {Computing in Science \& Engineering}\
  }\textbf {\bibinfo {volume} {9}},\ \bibinfo {pages} {90} (\bibinfo {year}
  {2007})}\BibitemShut {NoStop}%
\bibitem [{\citenamefont {Lewis}(2019)}]{Lewis:2019xzd}%
  \BibitemOpen
  \bibfield  {author} {\bibinfo {author} {\bibfnamefont {A.}~\bibnamefont
  {Lewis}},\ }\href@noop {} {\  (\bibinfo {year} {2019})},\ \Eprint
  {https://arxiv.org/abs/1910.13970} {arXiv:1910.13970} \BibitemShut {NoStop}%
\bibitem [{\citenamefont {{Scott}}(2012)}]{pippi}%
  \BibitemOpen
  \bibfield  {author} {\bibinfo {author} {\bibfnamefont {P.}~\bibnamefont
  {{Scott}}},\ }\href@noop {} {\bibfield  {journal} {\bibinfo  {journal}
  {\epjp}\ }\textbf {\bibinfo {volume} {127}},\ \bibinfo {pages} {138}
  (\bibinfo {year} {2012})},\ \Eprint {https://arxiv.org/abs/1206.2245}
  {arXiv:1206.2245} \BibitemShut {NoStop}%
\bibitem [{\citenamefont {Carron}(2013)}]{Carron_2013}%
  \BibitemOpen
  \bibfield  {author} {\bibinfo {author} {\bibfnamefont {J.}~\bibnamefont
  {Carron}},\ }\href {https://doi.org/10.1051/0004-6361/201220538} {\bibfield
  {journal} {\bibinfo  {journal} {Astronomy \& Astrophysics}\ }\textbf
  {\bibinfo {volume} {551}},\ \bibinfo {pages} {A88} (\bibinfo {year}
  {2013})}\BibitemShut {NoStop}%
\bibitem [{\citenamefont {Howlett}\ \emph {et~al.}(2017)\citenamefont
  {Howlett}, \citenamefont {Staveley-Smith},\ and\ \citenamefont
  {Blake}}]{Howlett_2016}%
  \BibitemOpen
  \bibfield  {author} {\bibinfo {author} {\bibfnamefont {C.}~\bibnamefont
  {Howlett}}, \bibinfo {author} {\bibfnamefont {L.}~\bibnamefont
  {Staveley-Smith}},\ and\ \bibinfo {author} {\bibfnamefont {C.}~\bibnamefont
  {Blake}},\ }\href {https://doi.org/10.1093/mnras/stw2466} {\bibfield
  {journal} {\bibinfo  {journal} {Monthly Notices of the Royal Astronomical
  Society}\ }\textbf {\bibinfo {volume} {464}},\ \bibinfo {pages} {2517–2544}
  (\bibinfo {year} {2017})}\BibitemShut {NoStop}%
\bibitem [{\citenamefont {Tegmark}(1997)}]{Tegmark_1997}%
  \BibitemOpen
  \bibfield  {author} {\bibinfo {author} {\bibfnamefont {M.}~\bibnamefont
  {Tegmark}},\ }\href {https://doi.org/10.1103/physrevlett.79.3806} {\bibfield
  {journal} {\bibinfo  {journal} {Physical Review Letters}\ }\textbf {\bibinfo
  {volume} {79}},\ \bibinfo {pages} {3806–3809} (\bibinfo {year}
  {1997})}\BibitemShut {NoStop}%
\bibitem [{\citenamefont {Hinton}\ \emph {et~al.}(2017)\citenamefont {Hinton},
  \citenamefont {Kazin}, \citenamefont {Davis}, \citenamefont {Blake},
  \citenamefont {Brough}, \citenamefont {Colless}, \citenamefont {Couch},
  \citenamefont {Drinkwater}, \citenamefont {Glazebrook}, \citenamefont
  {Jurek},\ and\ \citenamefont {et~al.}}]{Hinton_2016}%
  \BibitemOpen
  \bibfield  {author} {\bibinfo {author} {\bibfnamefont {S.~R.}\ \bibnamefont
  {Hinton}}, \emph {et~al.},\ }\href {https://doi.org/10.1093/mnras/stw2725}
  {\bibfield  {journal} {\bibinfo  {journal} {Monthly Notices of the Royal
  Astronomical Society}\ }\textbf {\bibinfo {volume} {464}},\ \bibinfo {pages}
  {4807–4822} (\bibinfo {year} {2017})}\BibitemShut {NoStop}%
\bibitem [{\citenamefont {Heavens}\ and\ \citenamefont
  {Sellentin}(2018)}]{Heavens:2018adv}%
  \BibitemOpen
  \bibfield  {author} {\bibinfo {author} {\bibfnamefont {A.~F.}\ \bibnamefont
  {Heavens}}\ and\ \bibinfo {author} {\bibfnamefont {E.}~\bibnamefont
  {Sellentin}},\ }\href {https://doi.org/10.1088/1475-7516/2018/04/047}
  {\bibfield  {journal} {\bibinfo  {journal} {JCAP}\ }\textbf {\bibinfo
  {volume} {04}},\ \bibinfo {pages} {047}},\ \Eprint
  {https://arxiv.org/abs/1802.09450} {arXiv:1802.09450} \BibitemShut {NoStop}%
\bibitem [{\citenamefont {{Athron}}\ \emph
  {et~al.}(2018{\natexlab{b}})\citenamefont {{Athron}}, \citenamefont
  {{Bal{\'a}zs}}, \citenamefont {{Bringmann}}, \citenamefont {{Buckley}},
  \citenamefont {{Chrz{\c a}szcz}}, \citenamefont {{Conrad}}, \citenamefont
  {{Cornell}}, \citenamefont {{Dal}}, \citenamefont {{Dickinson}},
  \citenamefont {{Edsj{\"o}}}, \citenamefont {{Farmer}}, \citenamefont
  {{Gonzalo}}, \citenamefont {{Jackson}}, \citenamefont {{Krislock}},
  \citenamefont {{Kvellestad}}, \citenamefont {{Lundberg}}, \citenamefont
  {{McKay}}, \citenamefont {{Mahmoudi}}, \citenamefont {{Martinez}},
  \citenamefont {{Putze}}, \citenamefont {{Raklev}}, \citenamefont {{Ripken}},
  \citenamefont {{Rogan}}, \citenamefont {{Saavedra}}, \citenamefont
  {{Savage}}, \citenamefont {{Scott}}, \citenamefont {{Seo}}, \citenamefont
  {{Serra}}, \citenamefont {{Weniger}}, \citenamefont {{White}},\ and\
  \citenamefont {{Wild}}}]{gambit_addendum}%
  \BibitemOpen
  \bibfield  {author} {\bibinfo {author} {\bibfnamefont {P.}~\bibnamefont
  {{Athron}}}, \emph {et~al.} (\bibinfo {collaboration} {\GB Collaboration}),\
  }\href@noop {} {\bibfield  {journal} {\bibinfo  {journal} {\epjc}\ }\textbf
  {\bibinfo {volume} {78}},\ \bibinfo {pages} {98} (\bibinfo {year}
  {2018}{\natexlab{b}})},\ \bibinfo {note} {addendum to \cite{gambit}},\
  \Eprint {https://arxiv.org/abs/1705.07908} {arXiv:1705.07908} \BibitemShut
  {NoStop}%
\end{thebibliography}%

\end{document}